\documentclass[preprint,3p,sort&compress,a4paper]{elsarticle}
\journal{Physica A}

\usepackage{graphicx}
\usepackage{subfig}
\usepackage{enumerate} 
\usepackage{amssymb}
\usepackage{amsthm}

\usepackage[]{hyperref}
\hypersetup{colorlinks}
\begin{document}

\begin{frontmatter}

\title{Dynamical analogy between epileptic seizures and seismogenic electromagnetic emissions by means of nonextensive statistical mechanics}

\author[1]{Konstantinos Eftaxias}
\author[2]{George Minadakis}
\author[3]{Stelios. M. Potirakis}
\author[4]{George Balasis}

\address[1]{Department of Physics, Section of Solid State Physics, University of Athens, Panepistimiopolis, GR 15784, Zografos, Athens, Greece}
\address[2]{Department of Electronic and Computer Engineering, Brunel University, Kingston Lane, Uxbridge, Middlesex, UB8 3PH, U.K.}
\address[3]{Department of Electronics, Technological Educational Institute of Piraeus, 250 Thivon \& P. Ralli, GR-12244, Aigaleo, Athens, Greece}
\address[4]{Institute for Space Applications and Remote Sensing, National Observatory of Athens, Metaxa and Vas. Pavlou St., Penteli, 15236 Athens, Greece}

\begin{abstract}
The field of study of complex systems considers that the dynamics of complex systems are founded on universal principles that may be used to describe a great variety of scientific and technological approaches of different types of natural, artificial, and social systems. Several authors have suggested that earthquake dynamics and neurodynamics can be analyzed within similar mathematical frameworks. Recently, authors have shown that a dynamical analogy supported by scale-free statistics exists between seizures and earthquakes, analysing populations of different seizures and earthquakes, respectively. The purpose of this paper is to suggest a shift in emphasis from the large to the small scale: our analyses focus on a single epileptic seizure generation and the activation of a single fault (earthquake) and not on the statistics of sequences of different seizures and earthquakes. We apply the concepts of the nonextensive statistical physics to support the suggestion that a dynamical analogy exists between the two different extreme events, seizures and earthquakes. We also investigate the existence of such an analogy by means of scale-free statistics (the Gutenberg-Richter distribution of event sizes and the distribution of the waiting time until the next event). The performed analysis confirms the existence of a dynamic analogy between earthquakes and seizures, which moreover follow the dynamics of magnetic storms and solar flares.
\end{abstract}


\end{frontmatter}

\section{Introduction}

Complexity nowadays is a frequently used yet poorly defined - at least quantitatively speaking - concept. It tries to embrace a great variety of scientific and technological approaches of all types of natural, artificial and social systems \cite{Tsallis2009}. In this direction several examples have been given providing the elements of such definition \cite{Bar-Yam1997,Picoli2007,Arcangelis2006,Kossobokov2000,Sornette2002,Abe2004,Fukuda2003,Peters2002}. When one considers a phenomenon that is ``complex'' refers to a system whose phenomenological laws, which describe the global behaviour of the system, are not necessarily directly related to the ``microscopic'' laws that regulate the evolution of its elementary parts \cite{Vicsek2002}. The field of study of complex systems considers that the dynamics of complex systems are founded on universal principles that may used to describe disparate problems \citep{Bar-Yam1997}. This is a basic reason for our interest in complexity \citep{Stanley1999,Stanley2000a,Sornette2002,Vicsek2001,Vicsek2002}. There is an apparent paradox in the above mentioned suggestion. How is it possible for a concept as multifaceted as complexity to serve as a unifying direction? There is a common factor in these seemingly diverse phenomena. The complex systems adopt a pattern of behaviour almost completely determined by the collective effects. They exhibit remarkable properties of self-organization and emergence of coherent structure over many scales. The main feature of collective behaviour is that an individual unit's action is dominated by the influence of its neighbours; the unit behaves differently from the way it would behave on its own, so that, all units simultaneously alter their behaviour to a common pattern. Thus, new features emerge as we move from one scale to another and the science of complexity is about revealing the principles that govern the ways in which new properties appear \cite{Vicsek2002}.

Complexity measures are generally performed to study pattern, structure and correlation of systems and processes. Although the definition of the term 'complexity' is not unique depending on the different contexts and different fields of application, its quantification has received considerable attention \cite[and references therein]{Telesca2011b}. In any case quantities such as: the disequilibrium,the randomness or uncertainty, the degree of order or organization, have played significant role on giving the notion of complexity \cite[and references therein]{Telesca2011b}. Complex systems seem to occur close, in some sense, to the frontier between order and disorder where most of their basic quantities exhibit nonexponential behaviours, very frequently power laws. It happens that the distributions and other relevant quantities that emerge naturally within the frame of nonextensive statistical mechanics are precisely of this type \cite{Tsallis2009}. Herein our analysis is mainly focused on the nonextensive Tsallis statistical mechanics, namely, a recently introduced nonextensive model for EQ dynamics \cite{Sotolongo2004} and Tsallis entropy. One universal footprint seen in many complex systems is self-affinity and the fractional power law relationship is a classic expression of a self-affine structure - the fractal. Note that the nonextensive model also leads to a fractional power law relationship \cite{Sarlis2010,Telesca2012mle}.

The contribution in this work enhances the suggestion that transferring ideas, methods and insights from investigations in hitherto disparate areas will cross-fertilize and lead to important new results. We stress that time-clustering methods are so powerful in describing so diverse natural as well as human phenomena. For instance, natural phenomena as well as human activities were found to share common time-clustering behaviour \cite{Telesca_1,Telesca_2,Telesca_3,Telesca_4,Telesca_5,Telesca_6,Telesca_7}. 

Epileptic seizures (ESs) and earthquakes (EQs) are complex phenomena, which have highly intricate cluster and hierarchical structures, spatial and temporal correlation with feedback, self-organization and connection diversity. Authors have suggested that dynamics of EQs and neurodynamics can be analyzed within similar mathematical frameworks \citep{Herz1995,Rundle2002}. Characteristically, driven systems of interconnected blocks with stick-slip friction capture the main features of EQ process. These models, in addition to simulating the aspects of EQs and frictional sliding, may also represent the dynamics of neurological networks \citep{Herz1995}. Hopfield \citep{Hopfield1994} proposed a model for a network of $N$ integrate-and-fire neurons. In this model, the dynamical equation of $k^{th}$ neuron, see equation 28 in \citep{Hopfield1994} is based on the Hodgekin-Huxley model for neurodynamics and represents the same kind of mean field limit that has been examined in connection with EQs \citep{Rundle2002}.

Recently, Osorio et al. \citep{Osorio2010} in a pioneering work have shown that a dynamical analogy supported by scale-free statistics exists between ESs and EQs. More precisely, the authors performed the analysis using: (i) a population of different EQs between 1984-2000 available in the Southern California Seismic Network catalogue, and (ii) a population of different ESs from continuous multiday voltage recordings directly from the brains of 60 human subjects with epilepsy undergoing surgical evaluation at the University of Kansas Medical Center between 1996 and 2000. In this work we investigate the existence of a dynamical analogy between ESs and EQs at the level of a single fault / seizure activation, namely, we examine whether a dynamical analogy may exist for the ways in which firing neurons / opening cracks organize themselves to produce {\it a single} ES / EQ. 

We perform the analysis mainly based on a recently introduced nonextensive model for EQ dynamics which leads to a Gutenberg-Richter type law for the relationship between frequency and magnitude of EQs \cite{Sotolongo2004,Silva2006}. We show that the populations of: (i) electric pulses included in a single ES, (ii) fracto-electromagnetic pulses rooted in the activation of a single fault, (iii) different ESs occurred in different human's brains, (iv) different EQs occurred in different faults in various seismic regions, follow the above mentioned nonextensive statistical law. The similarity is extended up to the laboratory seismicity. Importantly, the nonextensive model is also able to describe solar flares and magnetic storms, as well.  

Power-law correlations in both space and time are at least required in order to verify dynamical analogies between different catastrophic events. Hence, one may ask how the fluctuations included in a single EQ / ES precursory signal correlate in time. We investigate a possible temporal clustering focusing on a the existence of a common potential power-law distribution of burst lifetime (duration) in the populations of: (i) electric pulses included in a single ES, (ii) fracto-electromagnetic pulses rooted in the activation of a single fault, and (iii) acoustic pulses in laboratory. The result is positive. 

We claim that the performed analysis captured common principal laws behind the exciting variety of complex phenomena under study.

\section{Data Collection}

\subsection{Human and rat EEGs}

Electroencephalograms (EEG) are brain signals which provide insight concerning important characteristics of the brain activity and yield clues regarding the underlying neural dynamics.

Human EEG data offered by Andrzejak et al. \citep{Andrzejak2001} have been analysed. Two sets, denoted ``A'' and ``E'', respectively, each one of them containing 100 single-channel EEG segments of $23.6 sec$ duration, were employed for this study. Set ``A'' is comprised of EEGs of healthy volunteers. Set ``E'' contains seizure activities. The segments fulfil the criterion of stationarity \citep{Andrzejak2001}. 

EEG signals available from the on-line PhysioNET database (\url{http://physionet.org/pn6/chbmit/}), collected at the Children's Hospital Boston, have also been analyzed. This database consists of EEG recordings from pediatric subjects with intractable seizures. More details can be found in \cite{Goldberger2000}. In contrast to the data found in \citep{Andrzejak2001}, where set ``E'' contain exclusively seizure activities, these human EEG correspond to signals acquired throughout the whole evolution of the phenomenon (i.e., capturing the electric activity before, during and after the seizure). 

Moreover, adult Sprague-Dawley rats were used to study the epileptic seizures in EEG recordings \citep{Li2005,Kapiris2005}. A bicuculline i.p. injection was used to induce the rat epileptic seizures, while EEG signals were recorded. The seizure onset time was determined by visual identification of a clear electrographic discharge, and then looking backwards in the record for the earliest EEG changes from baseline associated with the seizure. The earliest EEG change is selected as the seizure onset time. Note that this data were mainly analyzed because they correspond to a controlled epilepsy, where the temporal stages (injection, pre-ictal, ictal and post-ictal), have been well justified for their origin. The exact time of the seizure provocation was also given.

\subsection{Pre-seismic EM emissions}

A question effortlessly arises whether there is an ``EM-seismograph'' available which can be used to monitor the evolution of a single fault activation process, in analogy to the EEG which is used to monitor the evolution of a single seizure activation process. Such a signal has been reported \cite{EftaxiasBook2012}. Fracture induced MHz-kHz electromagnetic (EM) fields allow a real-time monitoring of damage evolution in materials during mechanical loading. Crack propagation is the basic mechanism of material failure. EM emissions are produced by opening cracks. The radiated EM precursors are detectable both at a laboratory and geological scale \cite[and references therein]{Karamanos2006}. Especially, accumulated laboratory and field data suggest that the kHz EM activity, which is emerged in the tail of the observed EM radiation, is rooted in the fracture of strong entities distributed along the beleaguered by stresses fault sustaining the system \cite{Kapiris2004,Contoyiannis2005,Papadimitriou2008,Eftaxias2010uni}.

Herein, we mainly refer to the well documented seismogenic kHz EM activities associated with the: Athens (Greece) EQ ($M = 5.9$) that occurred on September 7, 1999 \citep{Eftaxias2001a,Kapiris2004,Contoyiannis2005,Karamanos2006,Eftaxias2007a,Papadimitriou2008,Kalimeri2008}.

\section{Fundamentals of nonextensive statistical mechanics}

Before presenting our arguments for the problem under study, we briefly introduce the nonextensive formalism to which we are referring to. If the physical system involves long-range interactions or long-range microscopic memory or (multi-)fractal boundary conditions, it can exhibit a quite anomalous thermodynamic behaviour, which might even be untraceable within Boltzmann-Gibbs (BG) statistical mechanics. To overcome at least some of these pathological situations, an entropic form $S_{q}$ has been proposed  by Tsallis  \citep{Tsallis1988,Tsallis2009}, which yields a generalization of standard statistical mechanics and thermodynamics. This entropy is defined as:

\begin{equation}
S_{q}=k\frac{1}{q-1}\left(1-\sum_{i=1}^{W}p_{i}^{q}\right),
\label{eq:t1}
\end{equation}

\noindent where, $p_{i}$ are the probabilities associated with the microscopic configurations, $W$ is their total number, and $k$ is a positive constant (from now on taken to be unity, without loss of generality). The value $q =1$ corresponds to the standard, extensive, BG statistics. 

A property which characterizes the above generalized entropic form is the following: if we have two independent systems $A$ and $B$ such that $p_{ij}^{A}= p_{i}^{A}+ p_{j}^{B}$, the Tsallis entropy satisfies 

\begin{equation}
S_{q}(A+B)=S_{q}(A)+S_{q}(B)+(1-q)S_{q}(A)S_{q}(B).
\label{eq:t2}
\end{equation}

The last term on the right side of the last equation brings the origin of nonextensivity of the resulting generalized statistical mechanics. The absolute value of $1-q$ indicates the degree of nonextensivity in a complex system. If $q > 1$, $= 1$ or $< 1$, $S_{q}$ is subextensive, extensive or superextensive, respectively. The index $q$ appears to characterise universality classes of nonadditivity, by phrasing this concept to what is done in the standard theory of critical phenomena. Within each class, one expects to find infinitely many dynamical systems \citep{Tsallis2009}.

Perhaps two of the most sound and rich examples of the dynamics of a complex system in crisis are the behaviour of brain / earth crust during the ES / EQ generation. A central property of their generation is the occurrence of large-scale collective behaviour with a very rich structure, resulting from repeated nonlinear interactions among the constituents, namely, firing neurons / opening cracks, of the system. Consequently, the nonextensive statistical mechanics \citep{Tsallis2009} is the appropriate framework in order to investigate the process of launch of the two shocks under study.

\section{Distinguishing the extreme events in terms of Tsallis entropy}

\begin{figure}[h]
\begin{center}
\includegraphics[width=1\textwidth]{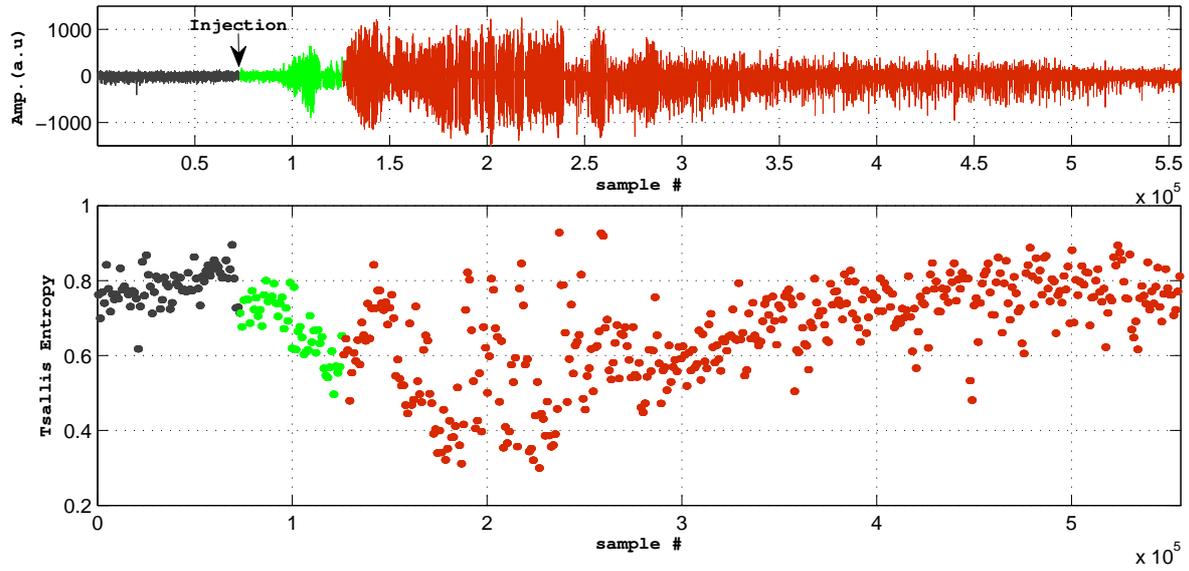}
\end{center}
\caption{The top graph depicts a rat EEG containing an evoked seizure. The arrow shows the time of injection. The black, green and red parts refer to the normal, pre-ictal and seizure epochs, correspondingly. The last stage includes the return to the normal state. The bottom graph shows the temporal evolution of Tsallis entropy, at non-overlapping windows of 1024 samples.}
\label{fig:en_rat}
\end{figure}

A basic step in the study of natural extreme events is to clearly distinguish them in the associated time series. The Tsallis entropy clearly discriminate the emergence of shocks in EEGs and kHz EM-seismograms by means of order of organization.  Indeed, a way to examine transient phenomena in a time-series is to analyze it into a sequence of distinct time windows of short duration and compute various measures of the degree of organisation content in each one of them. We employ here the Tsallis entropy in its symbolic form for word length $5$ and $q=1.7$ (see ref. \cite{Kalimeri2008}). A significant Tsallis entropy ``drop'' accompanies the appearance of two extreme events under study. More precisely:

Fig. \ref{fig:en_rat} shows the EEG of a rat along with its analysis in terms of Tsallis entropy. Three distinct phases in this time series were identified. The first (black) phase is the normal (healthy) state, before the applied injection which evokes the ES. It is followed by the pre-ictal (green) state which has higher organization in comparison to the organization of the healthy state. The first stage of the third (red) phase (ES) has a much higher degree of organization even in respect to those reported in the second phase. It is noted that the last stage includes the return to the normal state, while the Tsallis entropy gradually increases up to that of the healthy state. 

\begin{figure}[h]
\begin{center}
\includegraphics[width=1\textwidth]{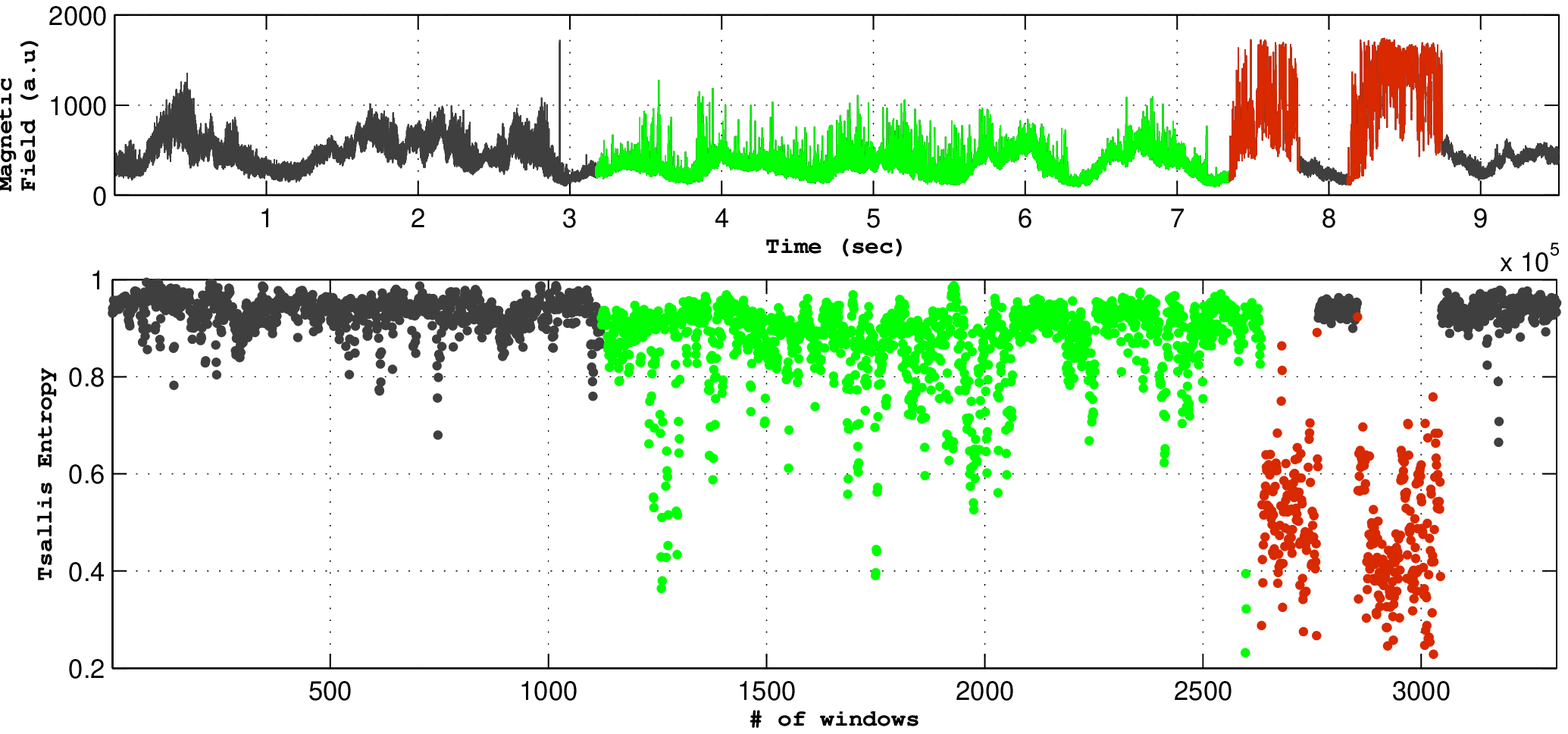}
\end{center}
\caption{The EM time series associated with the Athens EQ recorded by the 10 kHz magnetic sensor. The black part refers to the background noise, while the green and red parts refer to the two distinct epochs of the emerged seismic EM activity (see text). The bottom graph shows the temporal evolution of Tsallis entropy, using non-overlapping windows of 1024 samples.}
\label{fig:en_ath}
\end{figure}

Fig. \ref{fig:en_ath} refers to the kHz EM emission associated with the Athens EQ. Three distinct phases of time series were also identified. The first (black) part corresponds to the normal state (EM background), it is far from the EQ occurrence. It is followed by an epoch including a population of EM events sparsely distributed in time, which have higher organization in comparison to the organization of the EM background. The launch of the third epoch from the background is characterized by an ensemble of intense EM events, which have a much higher degree of organization even in respect to those reported in the second epoch. The last EM emission (comprised of two bursts) is thought to be due to the fracture of the family of large high-strength entities distributed along the fault sustaining the system \citep{Eftaxias2001a,Kapiris2004,Contoyiannis2005,Karamanos2006,Eftaxias2007a,Papadimitriou2008,Kalimeri2008}.

\begin{figure}[h]
\begin{center}
\includegraphics[width=1\textwidth]{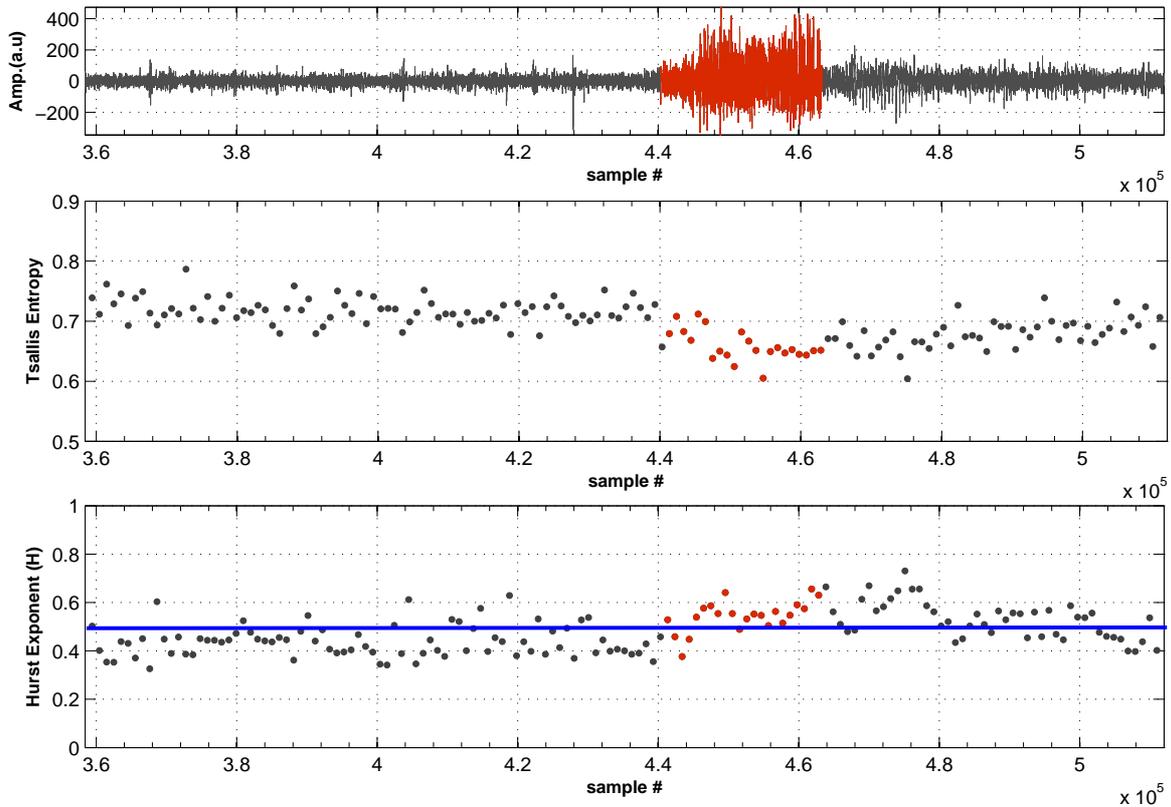}
\end{center}
\caption{The top graph depicts a human EEG that contains a seizure (red color), as given from the PhysioNET database \cite{Goldberger2000}. The bottom graph, shows the temporal evolution of Tsallis entropy using successive non-overlapping windows of 1024 samples each.}
\label{fig:en_hum}
\end{figure}

Note that both the examined extreme events (red epochs in Figs. \ref{fig:en_rat} and \ref{fig:en_ath}), have also been characterized by persistency, while the normal states (black epochs) have been characterized by antipersistency \cite{Kapiris2005}. The appearance of a high organization dynamics, which is simultaneously characterised by a positive feedback mechanism indicating a strong influence of excitation of an event on succeeding events, is consistent with the emergence of a catastrophic phenomenon and clearly discriminate the pathological from the healthy state. This suggestion is further verified by the analysis of human EEGs. Indeed, Fig. \ref{fig:en_hum} shows a human EEG along with its analysis in terms of Tsallis entropy and the Rescaled Range analysis (R/S), namely the Hurst exponent \cite{Hurst1951,Hurst1965}. The red part corresponds to the ictal phase, as has been annotated in the PhysioNET database \cite{Goldberger2000}. It is observed that the aforementioned scheme is also verified here. Characteristically, analysis in terms of Tsallis entropy, shows that the seizure part of the signal has a higher degree of organization than the rest of the signal. Additionally, as concerns the analysis in terms of the Hurst exponent, the seizure part presents persistent behaviour ($H>0.5$) in contrast to the rest of the signal. 

\clearpage

\begin{figure}[h]
\begin{center}
\includegraphics[width=1\textwidth]{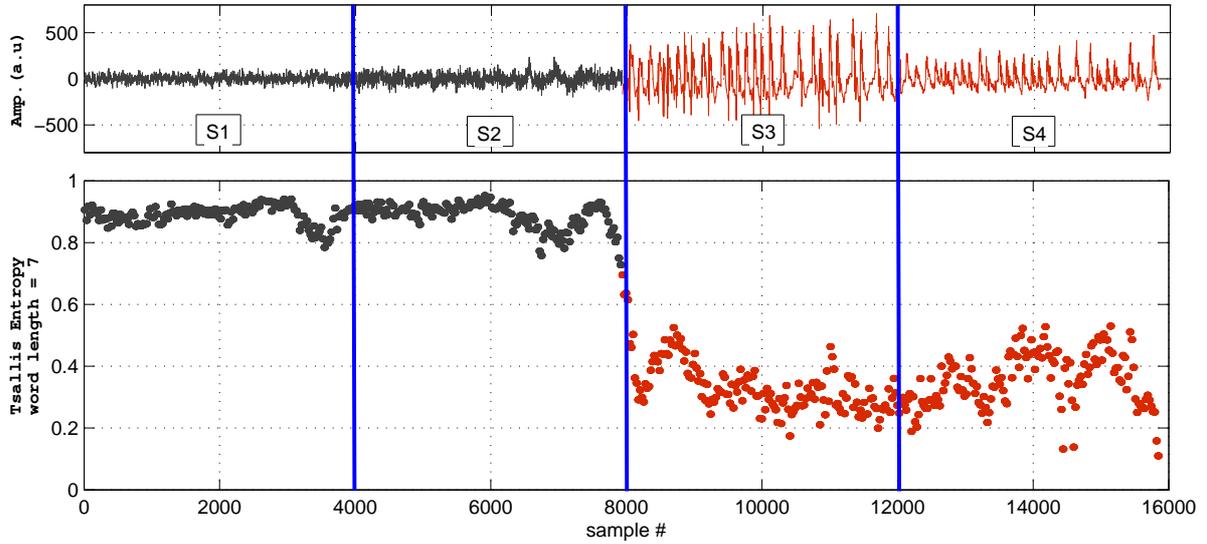}
\end{center}
\caption{The top graph, depicts a sequence of two healthy human EEGs and a sequence of two human epileptic seizures. The healthy EEGs are black coloured (S1,S2) whereas the patient EEGs are red coloured (S3,S4). }
\label{fig:humvshelthy}
\end{figure}

The higher degree of organization of the patients EEG compared to that of the healthy ones is further verified by analysing human EEGs. In Fig. \ref{fig:humvshelthy} we present two healthy (black) and two patients (red) EEGs included in the set A and E, respectively \citep{Andrzejak2001}. 

\begin{figure}[h]
\begin{center}
\subfloat {\includegraphics[width=0.5\textwidth]{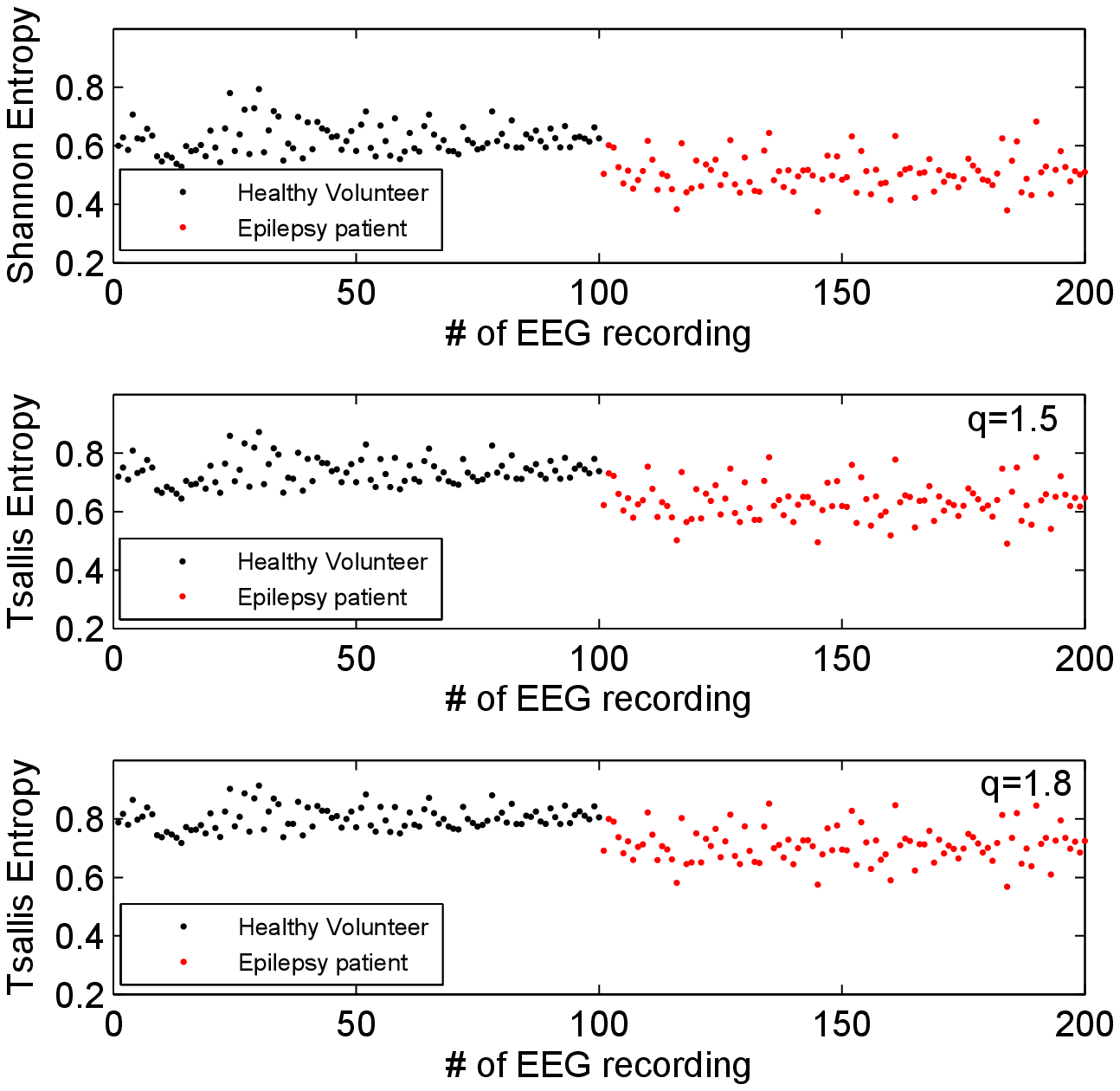}}
\subfloat {\includegraphics[width=0.5\textwidth]{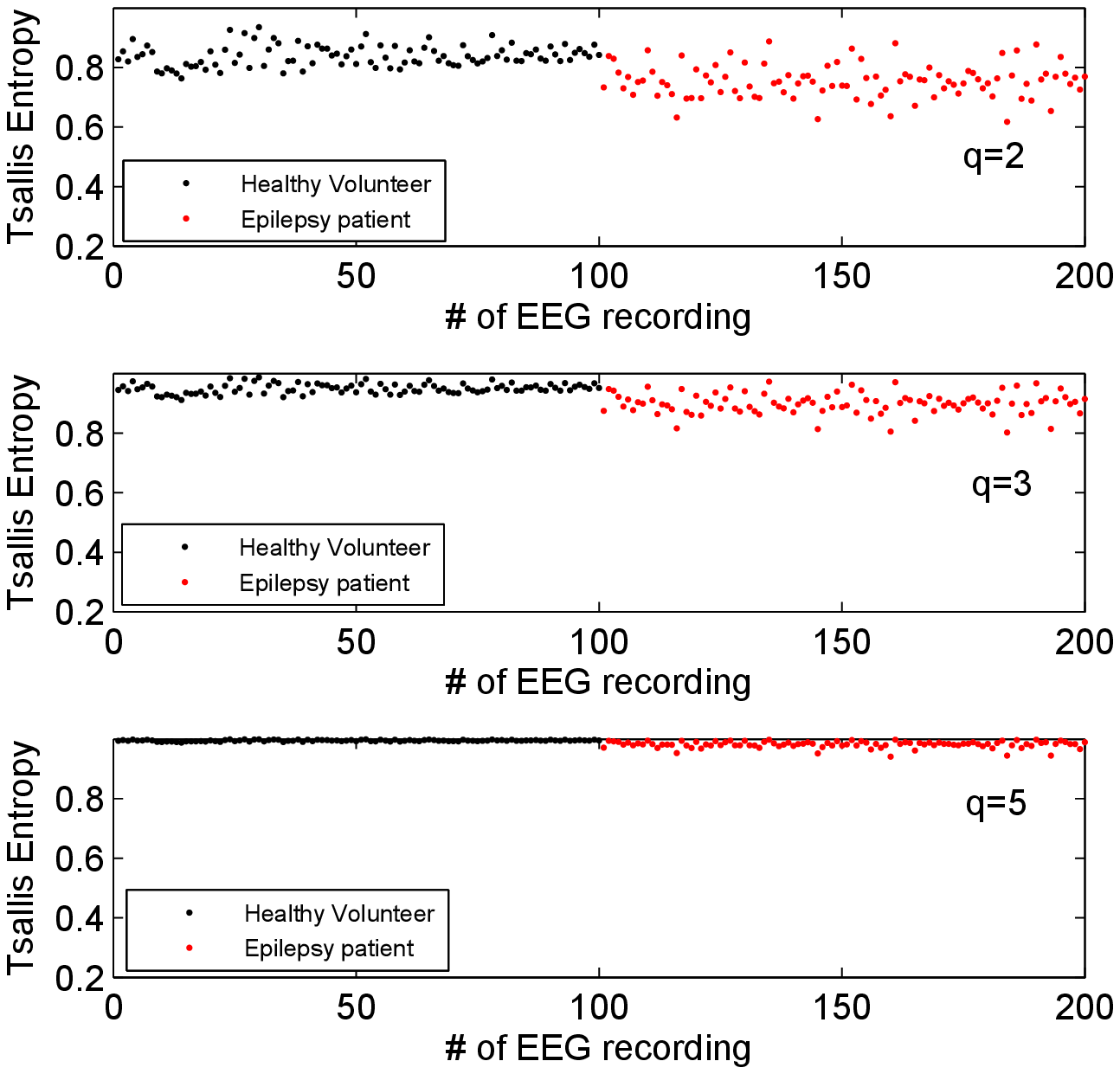}}
\end{center}
\caption{Shannon entropy (top-left), and Tsallis entropy of 100 healthy and 100 patient EEGs belonging to the ``A'' and ``B'' set, correspondingly \cite{Andrzejak2001}, for different values of $q$. The healthy EEGs are black colored while the patient EEGs are red coloured.}
\label{fig:en_100EEG}
\end{figure}

Fig. \ref{fig:en_100EEG}, shows the Tsallis entropies for 100 healthy and 100 patient human EEGs included in the sets A and E \citep{Andrzejak2001}. As expected, our results depend upon the Tsallis $q$ value. Figure \ref{fig:en_100EEG}, clearly illustrates the superiority of the $q$ values restricted in the range $1 < q < 2$ to magnify differences of the $S_{q}$. It is worth mentioning that the nonextensive $q$ parameter values which clearly quantify the temporal evolution of the complexity in the EEGs time series are in full agreement with the upper limit $q < 2$ obtained from several studies involving the Tsallis non-extensive framework \cite[and references therein]{Vilar2007}. Moreover, they are in harmony with an underlying sub-extensive system, $q > 1$, verifying the emergence of strong interactions in the brain, especially during the occurrence of an ES. However, it is expected that, for every specific system, better discrimination will be achieved with appropriates ranges of $q$ values \cite{Tsallis1988}. Thus, a challenge will be to estimate the appropriate value of $q$ which is associated with the generation of ESs. We attempt an estimation of the appropriate choice of the $q$ index in Fig. \ref{fig:en_100EEG}. We find that the appropriate $q$ index has a value of $q \sim 1.55$.

\clearpage
\section {Dynamical analogy in terms of nonextensive earthquake dynamics model}

A model for EQ dynamics has been recently introduced by Sotolongo-Costa and Posadas \citep{Sotolongo2004}. The mechanism of triggering EQs is established through the combination of the irregularities of the fault planes, on one hand, and the fragments between them, on the other hand. The fragments size distribution function comes from a nonextensive Tsallis formulation, starting from first principles, i.e., a nonextensive formulation of the maximum entropy principle. This nonextensive approach leads to a Gutenberg-Richter type law for the magnitude distribution of EQs. Silva et al. \citep{Silva2006} have subsequently revised this model considering the current definition of the mean value, i.e., the so-called $q$-expectation value as follows, 

\begin{eqnarray}
\log \left[N\left(>M\right)\right] = \log N+\left({2-q\over 1-q} \right)\log \left[1-\left({1-q\over 2-q} \right)\left({10^{2M} \over a^{2/3} } \right)\right] 
\label{eq:silva}
\end{eqnarray}

\noindent where, $N$ is the total number of EQs, $N(>M)$ the number of EQs with magnitude larger than $M$, and $M\approx \log \varepsilon$. $\alpha$ is the constant of proportionality between the EQ energy, $\varepsilon$, and the size of fragment.

Now, we examine whether Eq. (\ref{eq:silva}) describes sequences of: different EQs occurred in different faults in various seismic regions, fracto-EM events associated with the activation of a single fault, and electric events included in a single ES. 

(i) Eq. (\ref{eq:silva}) provides an excellent fit to different regional seismicities each of them including a population of earthquakes originated in different faults of the region under consideration. Importantly, the $q$-values are distributed within the range of ($1.6 - 1.8$) in various seismic regions studied \citep{Sotolongo2004,Silva2006,Matcharashvili2011,Telesca2010}. Recently Telesca \cite{Telesca2012mle} reported a Maximum Likelihood estimation of the nonextensive parameters of the EQ cumulative magnitude distribution.

(ii) The nonextensive formula (\ref{eq:silva}) describes the detected EM precursors associated with the activation of a single fault \citep{Papadimitriou2008,Eftaxias2001a}. We briefly focus on this point.

{\it The notion of ``EM-EQs''}: Supposing that the background (noise) level of the EM time series $A(t_{i})$ is $A_{noise}$, we regard as amplitude $A_{fem}$ of a candidate ``fracto-EM emission'' the difference $A_{fem}(t_{i})=A(t_{i})-A_{noise}$. We consider that a sequence of $k$ successively emerged ``fracto-EM emissions'' $A_{fem}(t_{i})$, $i=1,\ldots,k$ represents the EM energy released, $\varepsilon$, during the damage of a fragment. We shall refer to this as an ``electromagnetic earthquake'' (EM-EQ). Since the squared amplitude of the fracto-EM emissions is proportional to their power, the magnitude $M$ of the candidate EM-EQ is given by the relation:

\begin{equation}
M=\log\varepsilon \sim \log\left(\sum\left [ A_{fem}(t_{i})\right]^{2}\right),
\end{equation}

It has been shown that Eq. (\ref{eq:silva}) fits to the sequence of ``EM-EQs'' associated with the activation of a single fault, for example, Athens and L'Aquila EQs \citep{Papadimitriou2008,Kalimeri2008,Eftaxias2009,Eftaxias2010}. Herein, $N(>M)$ the number of ``EM-EQs'' with magnitude larger than $M$, and $\alpha$ the constant of proportionality between the EM energy released and the size of fragment. We note that, the best-fit parameter for these two precursors is given by $q \sim 1.80$, which is clearly in harmony with the exponents obtained in studies of regional seismicities \cite{Sotolongo2004,Silva2006,Papadimitriou2008,Telesca2010,Telesca2010b,Telesca2010c}. This result means that the activation of a single fault behaves as a "reduced image" of the regional seismicity as it was expected. Indeed, From the early work of Mandelbrot \cite{Mandelbrot1968}, the aspect of self-affine nature of faulting and fracture is widely documented from field observations, laboratory experiments, and studies of failure precursors on the small (laboratory) and large (EQ) scale. 

\begin{figure}[h]   
\begin{center}       
\subfloat []{\label{subfig:silva_1hum}\includegraphics[width=0.45\textwidth]{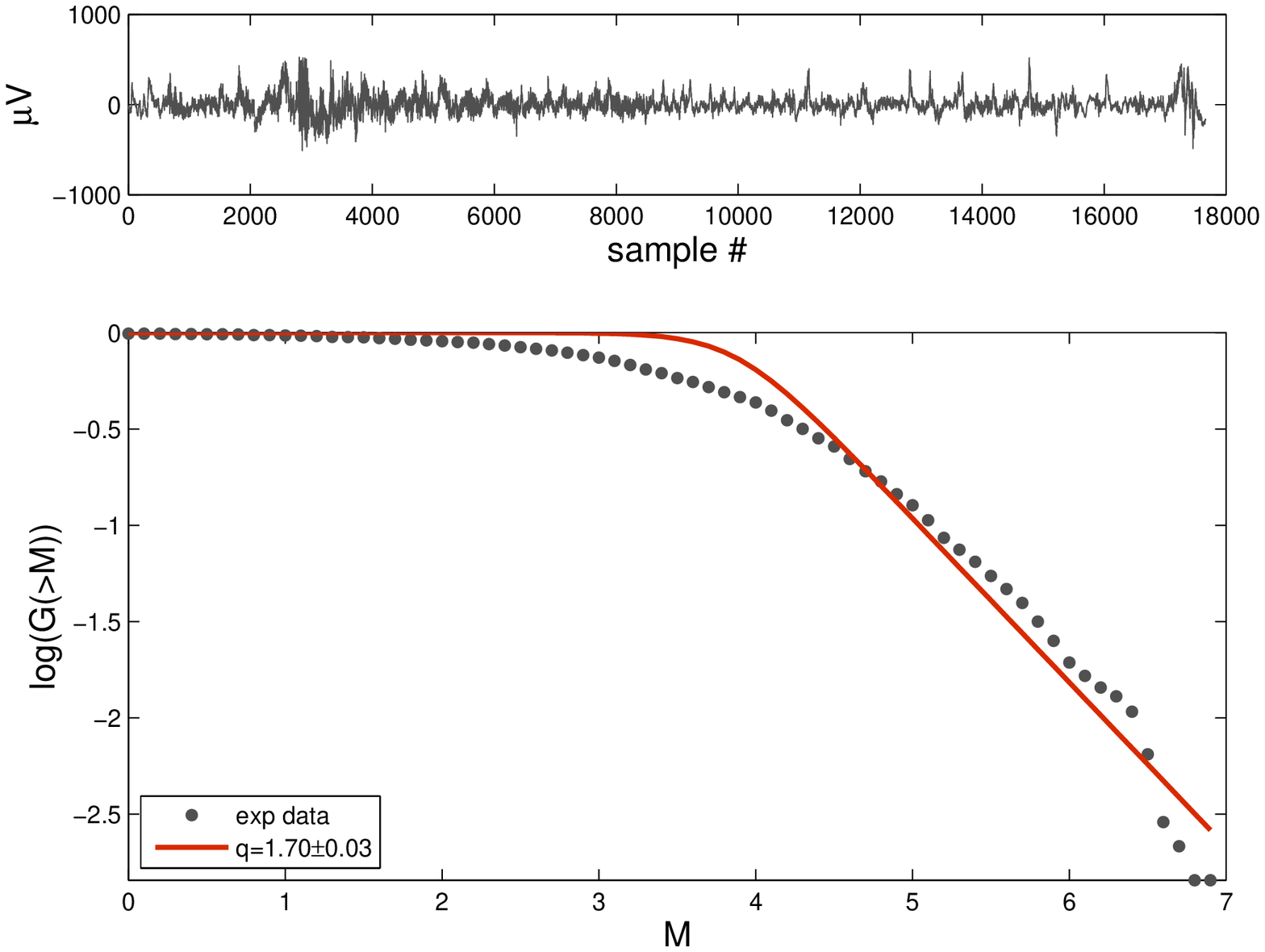}}
\subfloat []{\label{subfig:silva_2hum}\includegraphics[width=0.45\textwidth]{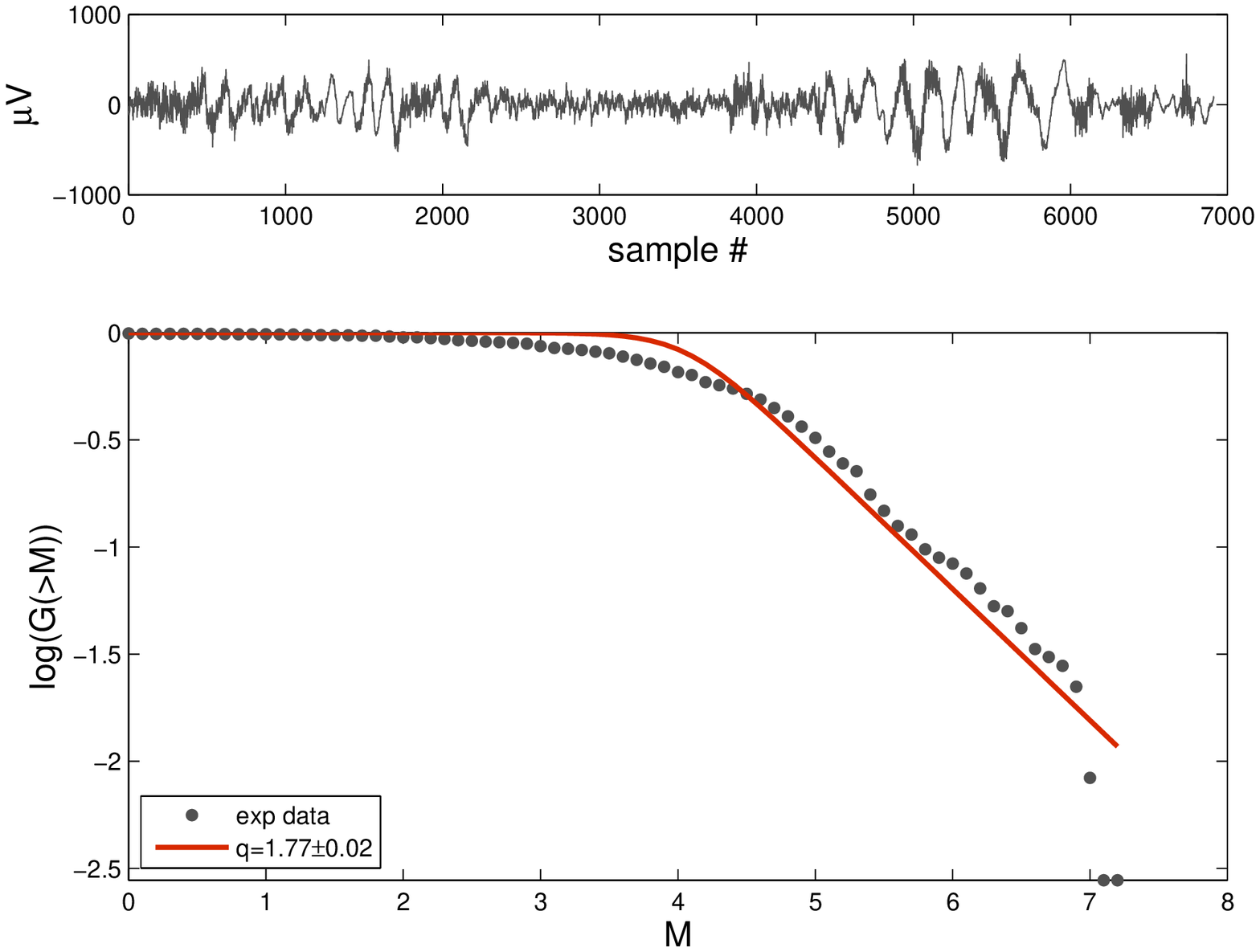}}

\subfloat []{\label{subfig:silva_3hum}\includegraphics[width=0.45\textwidth]{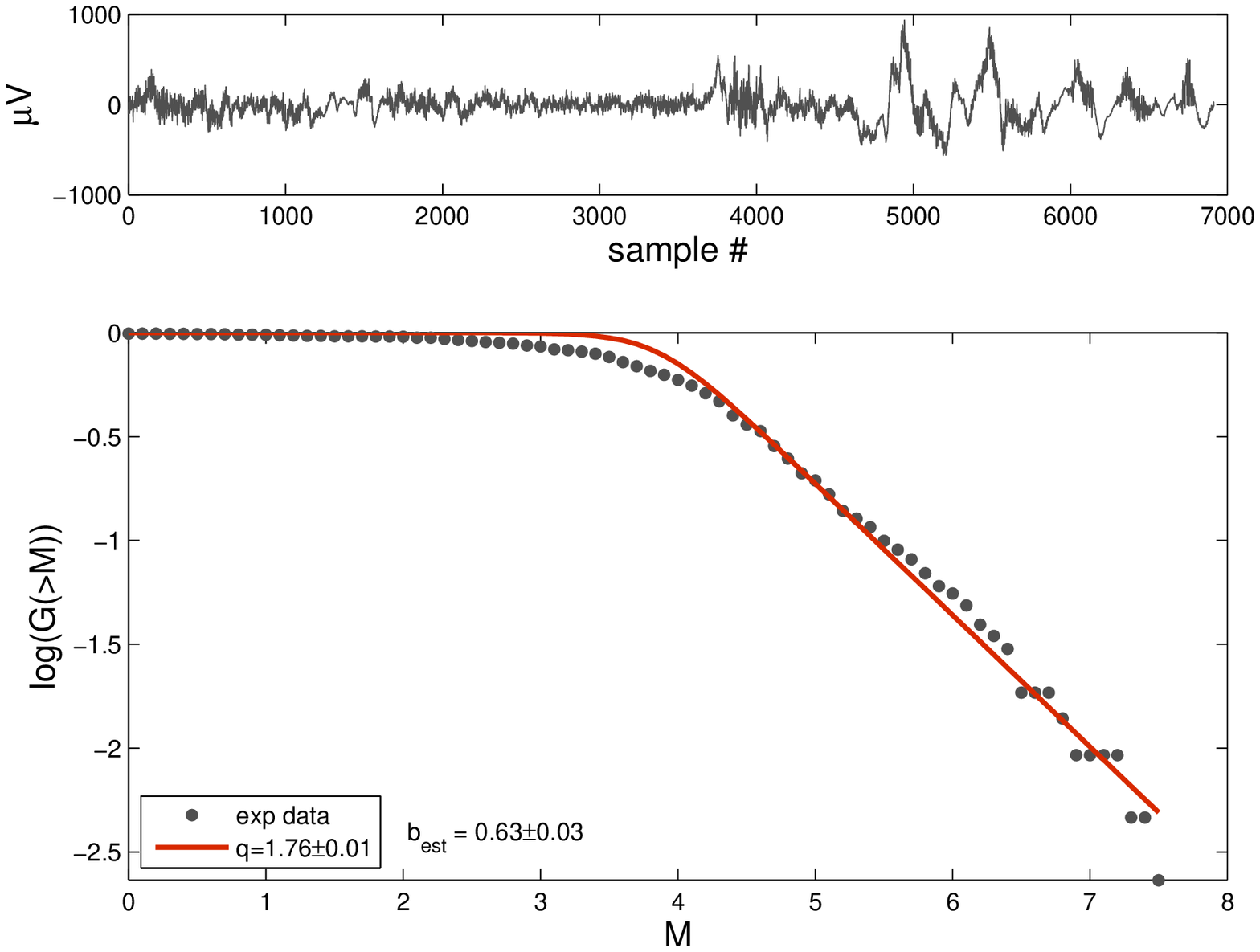}}
\subfloat []{\label{subfig:silva_4hum}\includegraphics[width=0.45\textwidth]{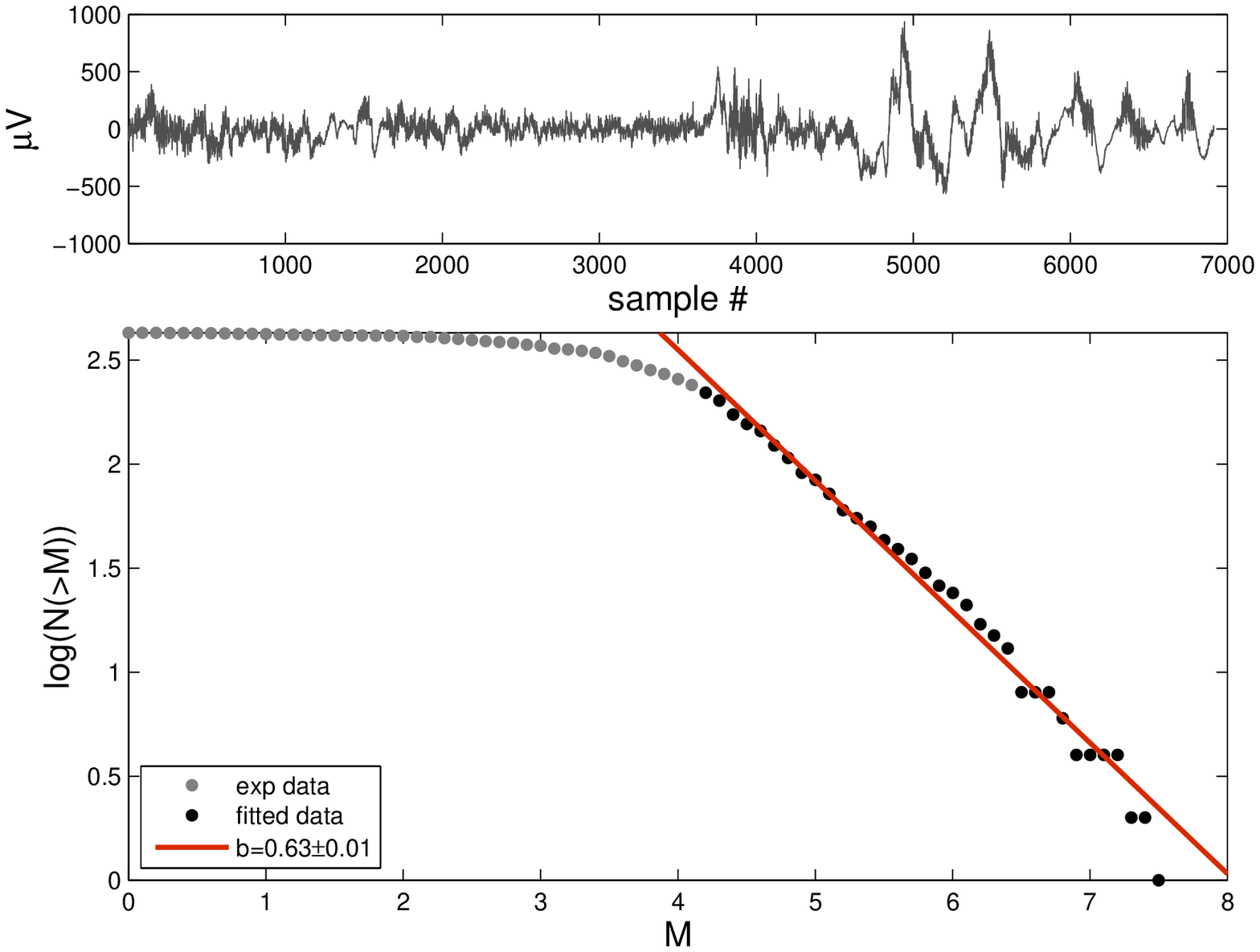}}
\end{center}   
\caption[]{(a)-(c) We used Eq. (\ref{eq:silva}) to fit the distribution of the magnitudes of electric events included in three single human epileptic seizures available from PhysioNET database \cite{Goldberger2000}. (d) Fitting of the data analyzed in Fig. \ref{subfig:silva_3hum} by the G-R law.}     
\label{fig:epi_silva1}
\end{figure}

(iii) We concentrate now on single human ESs. Figs. \ref{subfig:silva_1hum}, \ref{subfig:silva_2hum} and \ref{subfig:silva_3hum} show that the distribution of magnitudes of electric pulses included in three single ESs available from PhysioNET database \footnote{
\url{http://physionet.org/pn6/chbmit/chb01/chb01_04.edf}, Channel 14: F8-T8 /

\url{ http://physionet.org/pn6/chbmit/chb01/chb01_04.edf}, Channel 22: FT10-T8 / 

\url{http://physionet.org/pn6/chbmit/chb03/chb03_03.edf}, Channel 14: F8-T8.} \cite{Goldberger2000}, follow the nonextensive Eq. (\ref{eq:silva}) with $q=1.70\pm 0.03$, $q=1.77\pm 0.02$ and $q=1.76\pm 0.01$, respectively. Note that $G\left( >M \right)$ vs $M$ is depicted in Figs. \ref{subfig:silva_1hum}-\ref{subfig:silva_3hum} for convenience, where $G\left( >M \right)={N\left( >M \right)}/{{{N}_{tot.}}}\;$, with ${{N}_{tot.}}$ the total number of events. 

For reasons of completeness and in order to meet the analysis performed by Osorio et al. \citep{Osorio2010}, we also refer to the traditional Gutenberg-Richter law (G-R) which states that the number of EQs with magnitude greater than $M$  follows the relation: 

\begin{equation}
\log \left( N>M \right)=a-bM, 
\end{equation}
	
\noindent where $a$ and $b$ are constants describing the regional level of seismicity and relative size distribution of events respectively. 

We note that Sarlis et al. \cite{Sarlis2010} have shown that above some magnitude threshold Eq. \ref{eq:silva} reduces to the G-R law with 
\begin{equation}
b=2 \times \left( \frac{2-q}{q-1} \right)
\label{eq:sarlis}
\end{equation}

The experimental data shown in Fig. \ref{subfig:silva_3hum} were also directly fitted to G-R law for events with magnitude larger than $4.2$, yielding $b=0.63\pm 0.01$ (Fig. \ref{subfig:silva_4hum}). Note that the $q-$value resulting from by fitting Eq. (\ref{eq:silva}) to the specific data (Fig.\ref{subfig:silva_3hum}, $q=1.76\pm 0.01$) leads to an estimated, by Eq. (\ref{eq:sarlis}), ${{b}_{est}}=0.63\pm 0.03$ which coincides with that deduced via the direct use of the G-R law.

Figs. \ref{subfig:silva_1hum}-\ref{subfig:silva_4hum} show a superiority of the nonextensive Eq. (\ref{eq:silva}) to fit the data compared to the G-R law. There is always a clear drop off away from the linear trend of G-R law toward smaller magnitude values either we refer to natural or laboratory seismicity; this is also observed for the analyzed ES data. On the contrary, the nonextensive Eq. (\ref{eq:silva}), as can be seen from Figs. \ref{subfig:silva_1hum}-\ref{subfig:silva_3hum}, is capable of describing the magnitude distribution of electric pulses included in a single ES in the whole magnitude range. We emphasize that the aforementioned advantage of Eq. (\ref{eq:silva}) in respect to the G-R law is also observed in the description of natural seismicity \cite{Sotolongo2004,Silva2006}. Hence, Eq. (\ref{eq:silva}) provides the same physics for all the scales. 

\begin{figure}[h] 
\begin{center}
\subfloat []{\label{subfig:7a}\includegraphics[width=0.44\textwidth]{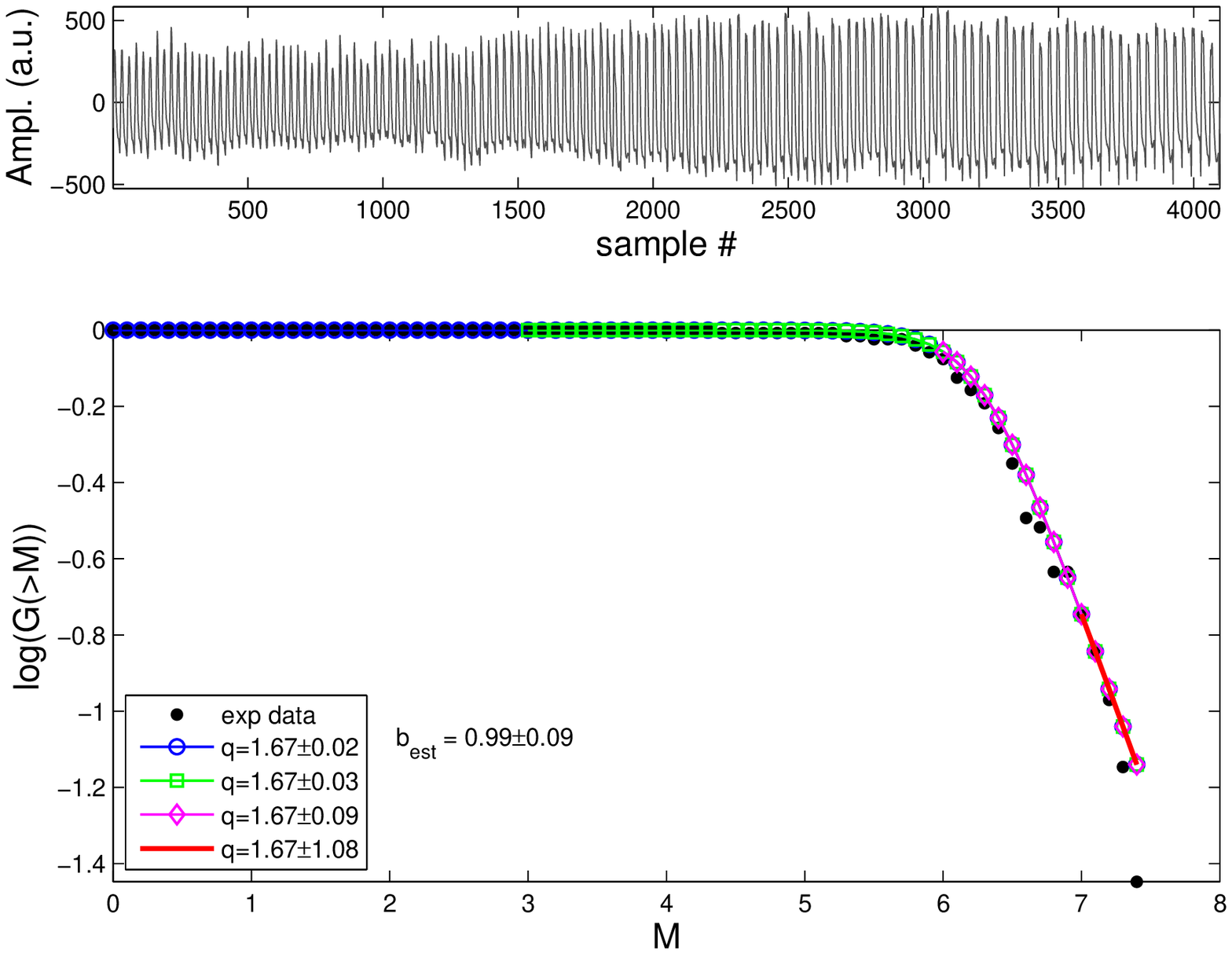}}
\subfloat []{\label{subfig:7b}\includegraphics[width=0.44\textwidth]{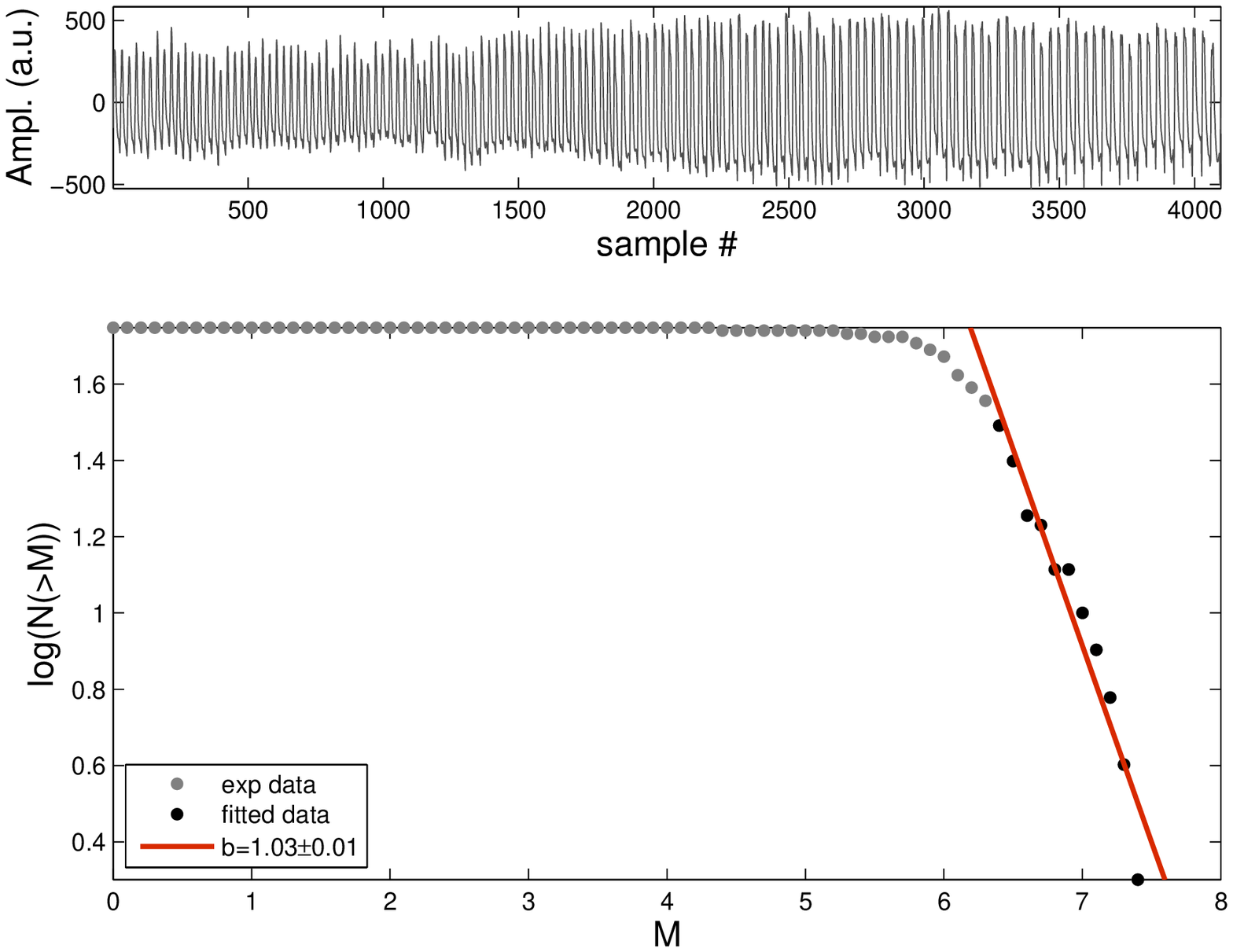}}

\subfloat []{\label{subfig:7c}\includegraphics[width=0.44\textwidth]{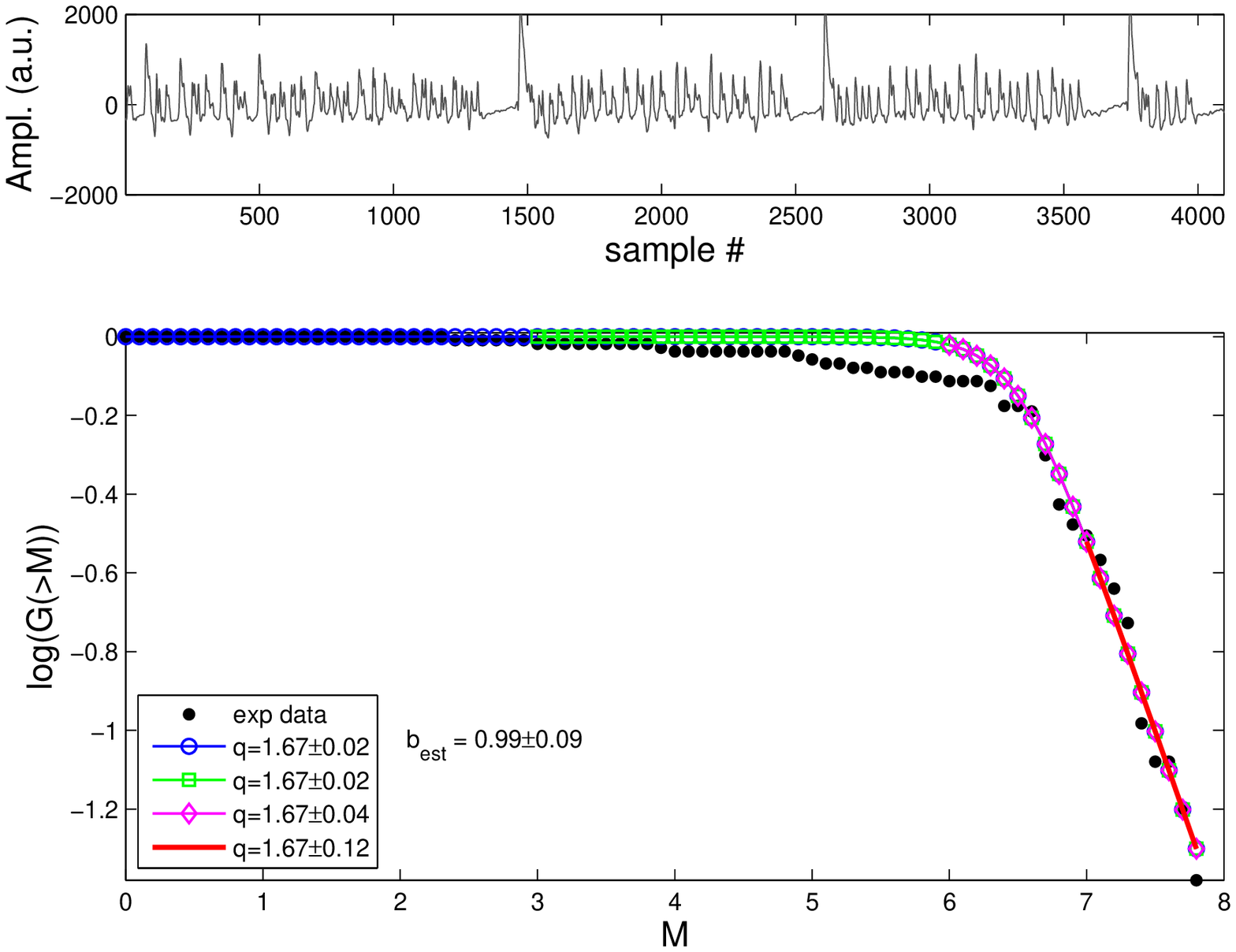}}
\subfloat []{\label{subfig:7d}\includegraphics[width=0.44\textwidth]{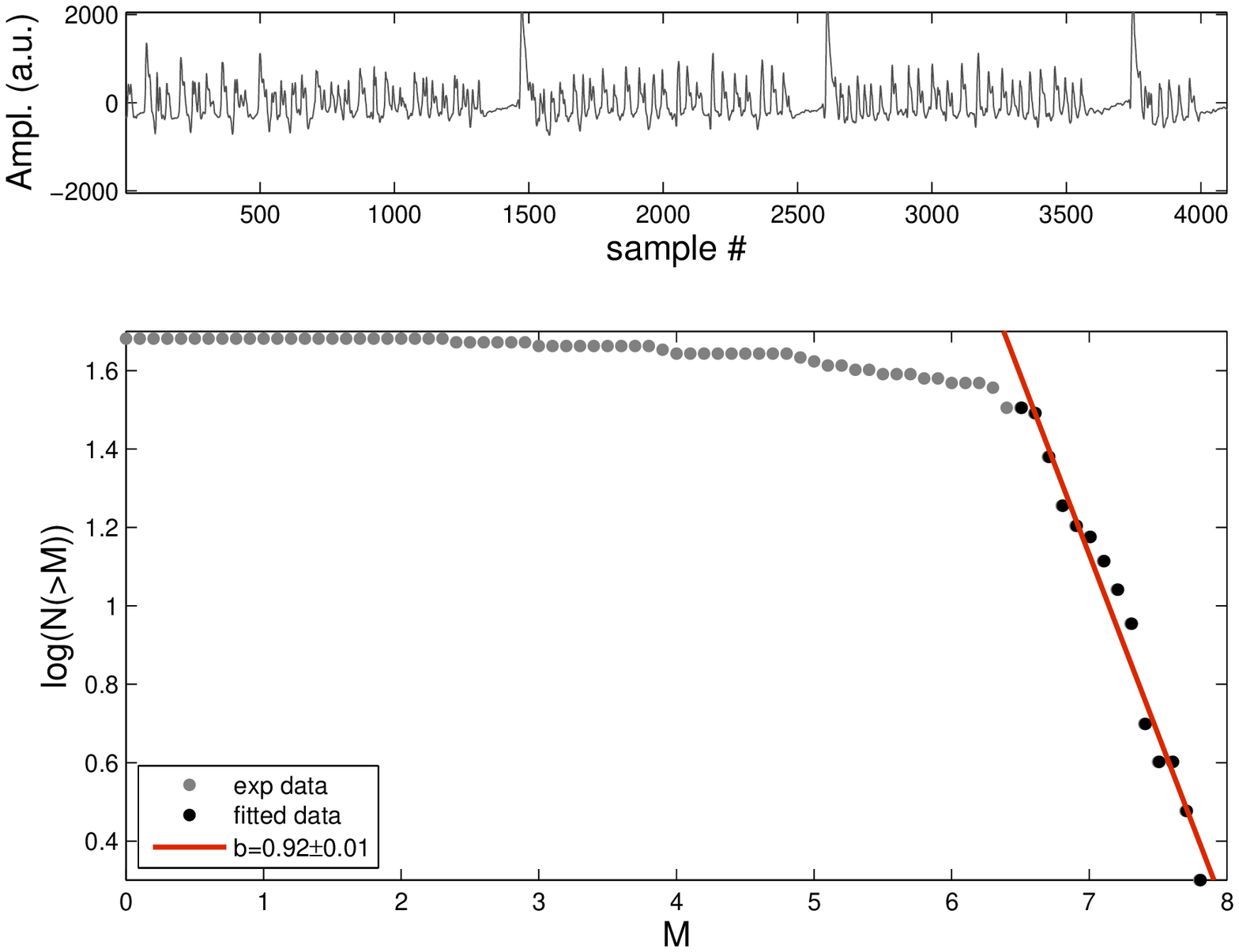}}

\subfloat []{\label{subfig:7e}\includegraphics[width=0.44\textwidth]{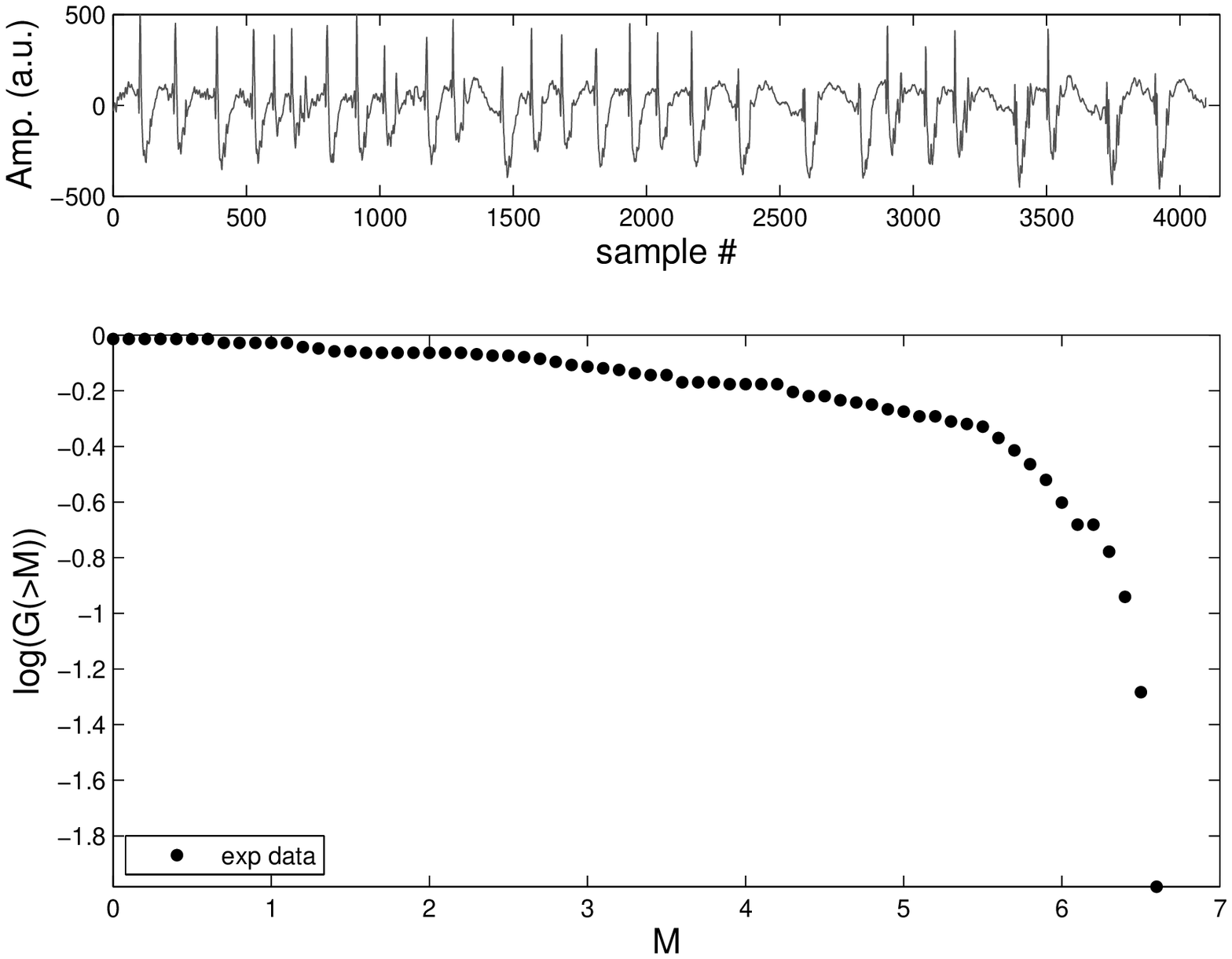}}
\subfloat []{\label{subfig:7f}\includegraphics[width=0.44\textwidth]{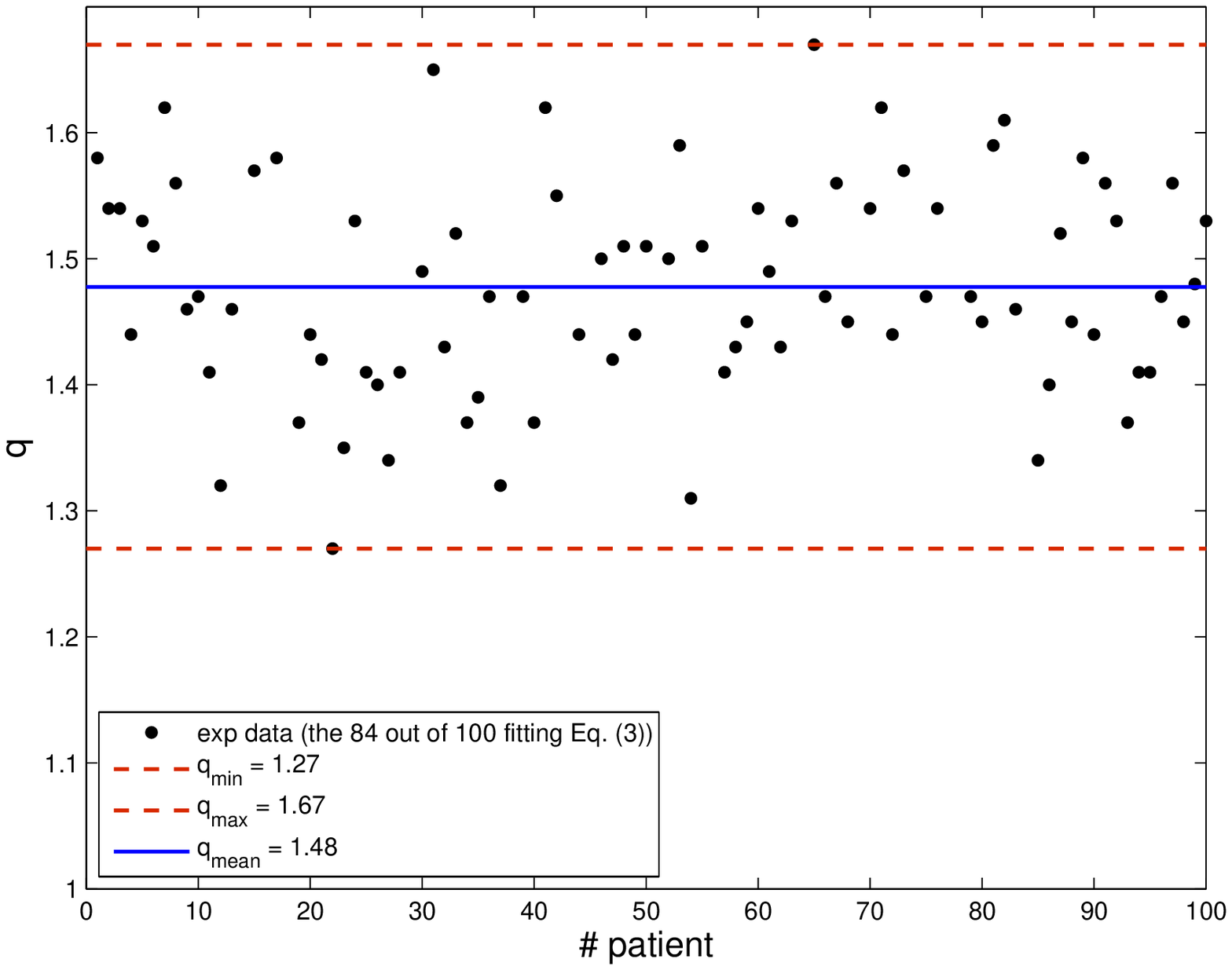}}

\end{center}
\caption[]{(a) The distribution of the magnitudes of electric events included in a single human seizure included in the set "E" \cite{Andrzejak2001} showing perfect agreement to the model expressed by Eq. (\ref{eq:silva}), (b) Fitting of the data analyzed in Fig. \ref{subfig:7a} by the G-R law. (c) Experimental data from a single human seizure included in the set "E" showing tolerable agreement to the model expressed by Eq. (\ref{eq:silva}), (d) Fitting of the data analyzed in Fig. \ref{subfig:7c} by the G-R law. (e) Experimental data not following either Eq. (\ref{eq:silva}) or G-R law (f) The values obtained by each one of the 84 out of 100 seizures belonging to the "E" set which could be fitted by Eq. (\ref{eq:silva}).}
\label{fig:epi_silva2}
\end{figure}

Now we refer to the human ESs included in the "E" \footnote{\url{http://epileptologie-bonn.de/cms/upload/workgroup/lehnertz/S.zip}} set \cite{Andrzejak2001}. The analysis of the 100 ESs of set E reveals that the vast majority of them (84\% of the analyzed ESs) can be fitted by Eq. (\ref{eq:silva}) yielding $q\in \left[ 1.27,1.67 \right]$ (Fig \ref{subfig:7f}), while the rest of them seem to follow quite different statistics than the one described by either Eq. (\ref{eq:silva}) or the G-R law (e.g., Fig. \ref{subfig:7e}). Among the successfully fitted cases, two different groups should be discriminated. In the first one, the distribution of electrical pulses included in a single ES is nicely described by the Eq. (\ref{eq:silva}) at all scales; Fig. \ref{subfig:7a} shows such an example. In the second group, the data can not be described by Eq. (\ref{eq:silva}) at all scales; the fit is poor lower than a magnitude threshold value (refer to the deviation that the experimental data show from the estimated curve in the "knee" region in Fig. \ref{subfig:7c}). However, attempting a series of fittings by Eq. (\ref{eq:silva}) using a different threshold values each time, always leads to the same estimated $q-$value, although with different accuracies (Fig. \ref{subfig:7c}). Note that the $q-$values resulting from the fitting of Eq. (\ref{eq:silva}) on the experimental data (for both Fig. \ref{subfig:7a} and Fig. \ref{subfig:7c} we obtain $q=1.67\pm 0.02$) lead to estimated $b-$values, by Eq. (\ref{eq:sarlis}), ${{b}_{est}}=0.99\pm 0.09$. The close agreement of the estimated and directly calculated $b-$values for the two different ES examples analyzed, $b=1.03\pm 0.01$ (Fig. \ref{subfig:7b}) and $b=0.92\pm 0.01$ (Fig. \ref{subfig:7d}), respectively, verifies that the nonextensive model of Eq. (\ref{eq:silva}) can successfully describe the analyzed ES data. 

The observed differentiation of the behaviour of SEs belonging to the revealed two different groups is not unexpected. It has been emphasized the great variability in the morphology and mode occurrence of seizures, and the great variability of non-ictal EEG patterns \cite{Khan2003}. Note that the 100 ESs of set E come from different patients and different recording regions of the brain exhibiting ictal activity. Therefore, the observed differentiation in fitting Eq. (\ref{eq:silva}) may be a reflection of the variability in the morphology of the analyzed ESs. Note that this hypothesis is further enhanced by the observed variety of organization among the different signals of set E (Fig. \ref{fig:en_100EEG}). Moreover, although the level of experimental noise is not given for the signals of set E, it is expected that different signals (from different patients / electrodes) may be characterized by different signal to noise ratios (SNRs). For the signals with lower SNR it is likely that the validity of events of magnitude lower than a specific threshold is questionable.

\clearpage

\clearpage
The above presented  results support the suggestion that a dynamical analogy in terms of nonextensive statistical mechanics exists between ES and EQ generation.  Though intriguing to some extent, the above mentioned results suggest that a more exhaustive study of the aforementioned biological and geophysical shocks in terms of nonextensive statistics is needed to give a deeper interpretation of their generation.

\section{Dynamical analogy by means of ``Gutenberg-Richter scaling law''}

In terms of energy the G-R law states that, the probability density function of having an EQ energy $E$ is denoted by the power-law $P(E)\sim E^{-B}$ where $B \sim 1.4-1.6$ \cite{Gutenberg1954}. The probability of an ES in a population of different events having energy $E$ is proportional to $E^{-B}$, where $B \sim 1.5-1.7$ \citep{Osorio2010}. 

\begin{figure}[h]
\begin{center}
\subfloat []{\label{subfig:endist_rat}\includegraphics[width=0.5\textwidth]{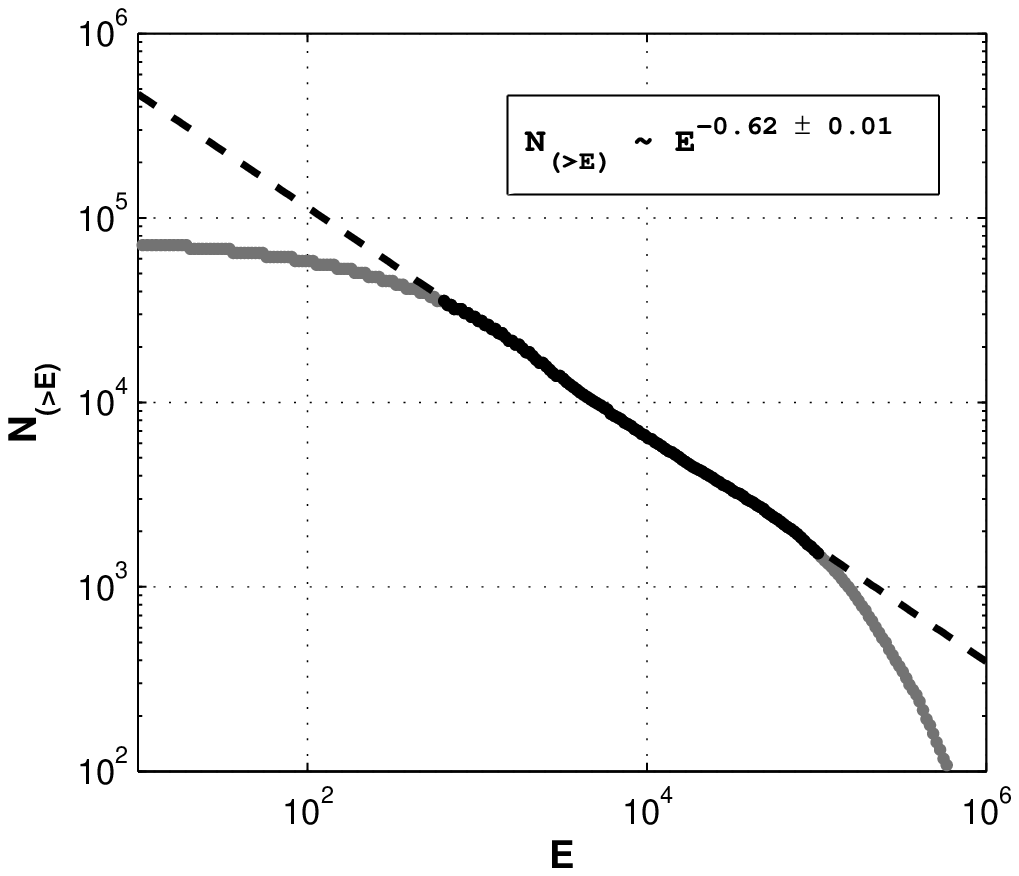}}
\subfloat []{\label{subfig:endist_hum}\includegraphics[width=0.5\textwidth]{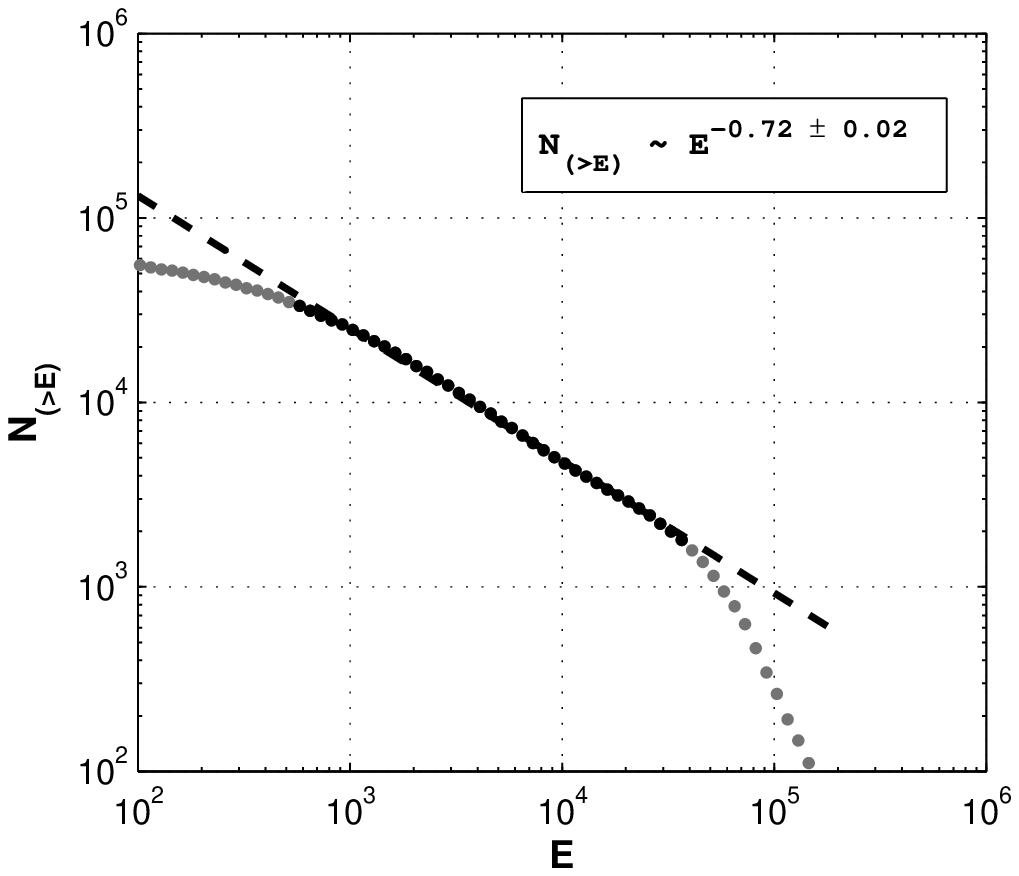}}
\end{center}
\caption{(a) The distribution of energies, $E$, of the electrical pulses included in the rat seizure depicted in Fig. \ref{fig:en_rat} follows the power-law $N(>E) \sim E^{-0.62}$. (b) The distribution of energies, $E$, of the electrical pulses included in a human ES, depicted in Fig. \ref{fig:en_hum}, follows the power-law $N(>E) \sim E^{-0.72}$.}
\label{fig:endists}
\end{figure}

Herein, we examine whether the sequences of electrical pulses included in a single ES follow a power-law $P(E)\sim E^{-B}$, with a compatible exponent. Fig. \ref{subfig:endist_rat}, shows that the energies, $E$, of the electrical pulses included in the single rat seizure depicted in Fig. \ref{fig:en_rat} follows the power-law $N(>E) \sim E^{-0.62}$, or equivalently, the power-law $N(E) \sim E^{-1.62}$. Fig. \ref{subfig:endist_hum}, depicts that the energies, $E$, of the electrical pulses included in a human ESs follows the power-law $N(>E) \sim E^{-0.72}$, or equivalently, the power-law $N(E) \sim E^{-1.72}$.

As regards the case of the kHz EM precursor associated with the Athens EQ (Fig. \ref{fig:en_ath}, upper panel), it has been shown that the cumulative number $N(>A)$ of pre-seismic EM pulses having amplitudes larger than $A$ follows the power-law $N(>A)\sim A^{-0.62}$ \citep{Kapiris2004b}. Thus, the probability of an EM-pulse having energy $E$ is proportional to $E^{-1.31}$ \citep{Eftaxias2004}.

The above mentioned results indicate that sequences of: (i) fracto-EM-pulses included in a single EM-precursor associated with the activation of a single fault, (ii) electric pulses included in a single ES, (ii) different EQs occurred in various areas included many faults, and (iv) different ESs, follow the power-law $P(E)\sim E^{-B}$ with rather compatible $B$-exponents. Notice, in general, differences in constituting elements (organic vs inorganic), in scale, and in other properties between the earth and brain may account for dissimilarities in the values of exponents \citep{Osorio2010}. Thus, the reported dynamical analogy in \citep{Osorio2010} between ESs and EQs by means of energy in populations of different ESs and EQs is extended to smaller scale, namely in the activation of a single ES / EQ. 

Importantly, the aforementioned dynamical analogy is extended up to the scale of laboratory seismicity. Acoustic / EM emission in rocks has been studied as a model of natural seismicity. Rabinovitch et al. \citep{Rabinovitch2001} have studied the fractal nature of EM radiation induced by rock fracture. Their analysis of the pre-fracture EM time series reveals that the cumulative distribution function of the amplitudes follows the power $N(>A)\sim A^{-0.62}$, and, consequently, the distribution function of the energies follows the power-law $P(E)\sim E^{-1.31}$, as it happens in the case of fracture EM radiation associated with the activation of a single natural fault. Petri et al. \citep{Petri1994} found a power-law scaling behaviour in the acoustic emission energy distribution with $B=1.3 \pm 0.1$. Houle and Sethna \citep{Houle1996} found that the crumpling of paper generates acoustic pulses with a power-law distribution in energy with $B=1.3-1.6$. On the other hand, Cowie et al. (1993) \cite{COWIE1993}; Sornette et al. (1994) \cite{Sornette1994}; Cowie et al. (1995) \cite{COWIE1995}, have developed a model of self-organized EQs occurring on self-organized faults. Their theoretical study suggests that the value should be $B=1.3$.

\section{Dynamical analogy in terms of waiting times}

Power-law correlations in both space and time are at least required in order to verify dynamical analogies between different catastrophic events. Hence, one can ask how the EM / electrical fluctuations included in a single EM precursor / ES correlate in time. We investigate the aforementioned temporal clustering in terms of burst lifetime (duration) focusing on a potential power-law distribution. We note that in \citep{Osorio2010}, the probability-density-function for intervened intervals $\tau$ was calculated for a population of different ESs. The statistic follows a power-law distribution $\sim 1/(\tau^{1+\beta})$, with $\beta\sim 0.5$ for interseizure intervals.

\begin{figure}[h]
\begin{center}
\subfloat []{\label{subfig:rat_dtimes}\includegraphics[width=0.33\textwidth]{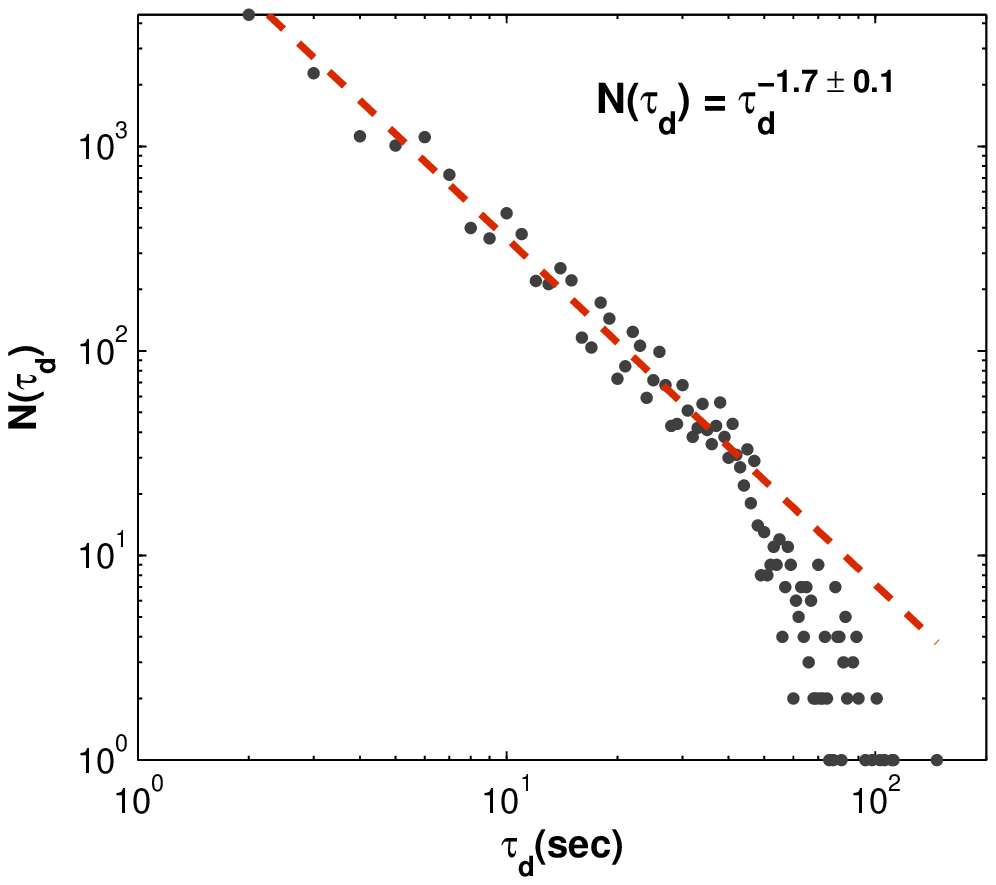}}
\subfloat []{\label{subfig:hum_dtimes}\includegraphics[width=0.33\textwidth]{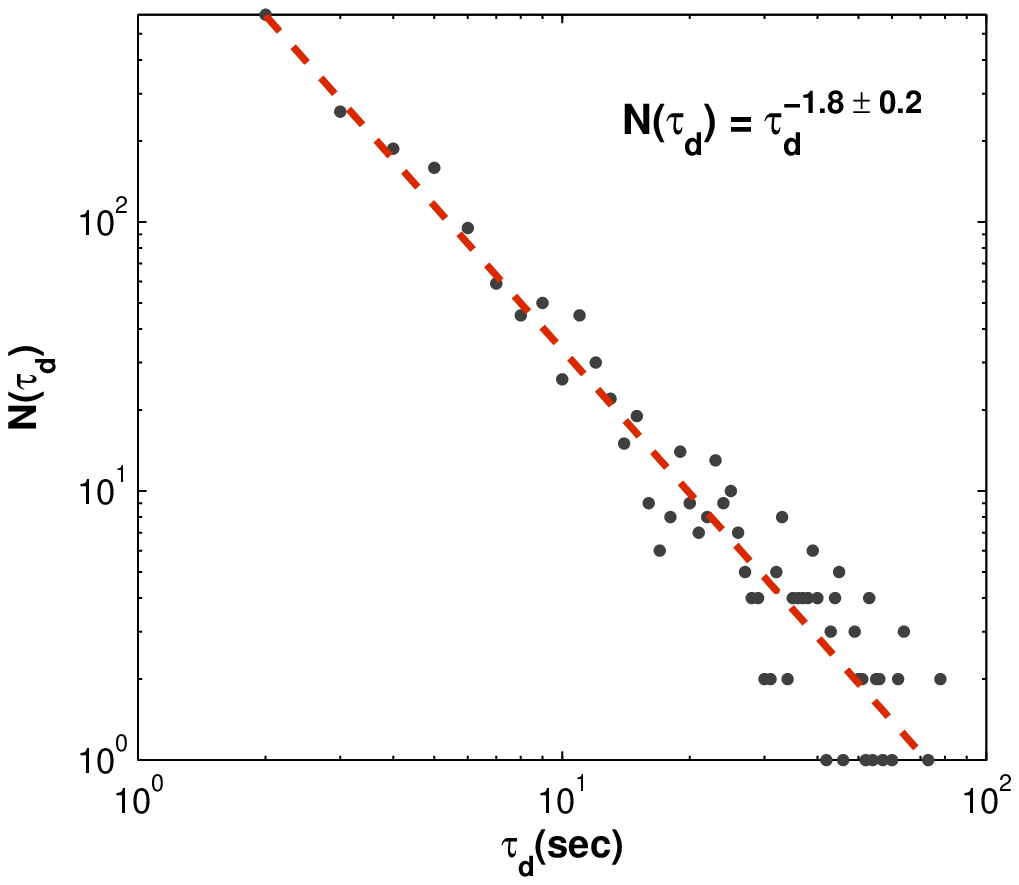}}
\subfloat []{\label{fig:EM_dtimes}\includegraphics[width=0.33\textwidth]{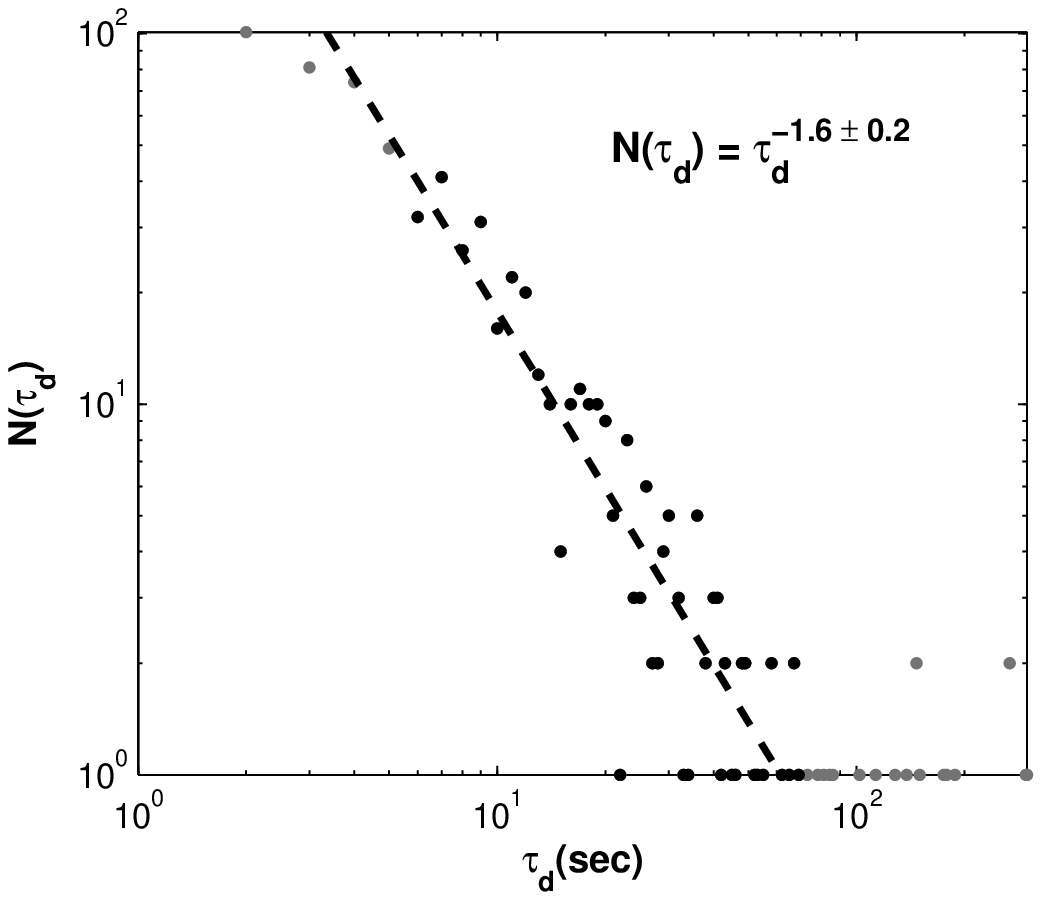}}

\end{center}
\caption{(a and b) Distributions of the lifetimes of the electric pulses included in a single rat (shown in Fig. \ref{fig:en_rat}) and human ES (shown in Fig. \ref{fig:en_hum}), correspondingly. (c) Distributions of the lifetimes of the EM pulses included in the precursory kHz EM activity associated to Athens earthquake. The time durations (livetime) of the emerged EM events, follows the power-law distribution $\sim 1 / \tau^{1.6}$.}
\end{figure}

Figs. \ref{subfig:rat_dtimes} and  \ref{subfig:hum_dtimes}, depict the distributions of the lifetimes of the electric pulses included in a single rat (shown in Fig. \ref{fig:en_rat}) and human ES (shown in Fig. \ref{fig:en_hum}), correspondingly. We observe that both time-series follow a power-law distribution $\sim 1 / {\tau_{w}}^{1.7}$ / $\sim 1 / {\tau_{w}}^{1.8}$, respectively. 

We refer to the pre-seismic EM activity associated with the Athens EQ (see Fig. \ref{fig:en_ath} upper panel). The analysis reveals that the durations (lifetimes) display a power-law distribution $\sim 1 / {\tau_{w}}^{1.6}$ (Fig. \ref{fig:EM_dtimes}). 

Note that Vespignani et al. (1995) \cite{Vespignani1995} measured an exponent $=1.6$ via acoustic signals from laboratory samples subjected to an external stress. The above mentioned findings further enhance the view that a dynamical analogy exists between seismic and epileptic events.

\section{Discussion \& Conclusions}

The field of study of complex systems considers that the dynamics of complex systems are founded on universal principles that may be used to describe disparate problems ranging from particle physics to economies of societies. This is a basic reason for our interest in complexity \citep{Bar-Yam1997,Stanley1999,Stanley2000a,Sornette2002,Vicsek2001,Vicsek2002,Picoli2007,Arcangelis2006,Kossobokov2000,Sornette2002,Abe2004,Fukuda2003,Peters2002}.

Authors have suggested that dynamics of EQs and neurodynamics can be analyzed within similar mathematical frameworks \cite{Hopfield1994,Herz1995,Rundle2002}. The present paper reports the first indication that a dynamical analogy exists between the generation of a single epileptic seizure and a single earthquake activation. The results of this work were built on the concepts of nonextensive statistical mechanics. More precisely, a recent nonextensive model for earthquake dynamics leads to a Gutenberg-Richter type law for the magnitude distribution of EQs (formula \ref{eq:silva}). We showed that the populations of fracto-EM-pulses rooted in the fracture of the backbone of strong entities distributed along a single fault sustaining the system, and electric pulses included in a single epileptic seizure follow the above mentioned nonextensive distribution with compatible nonextensive $q$-parameter.

Our analysis also revealed common ``pathological symptoms'' of a transition to the generation of seemingly different two extreme events under study. The generation of a single epileptic seizure/ earthquake is accompanied by the appearance of: (i) significant lower non-extensive Tsallis entropy, namely, higher organization, and (ii) persistency, i.e., a mechanism which is characterized by a positive feedback, thus leading the complex systems out of equilibrium.

The nonextensive formula (formula \ref{eq:silva}) also describes the magnitude distribution of solar flares and magnetic storms with well-matched nonextensive $q$-parameter. The two last extreme events also are accompanied by the appearance of the above mentioned 
``pathological symptoms'', namely, higher organization and persistency \cite{Kapiris2004,Kapiris2005,Karamanos2006,Contoyiannis2005,EftaxiasBook2012,Balasis2008,Balasis2011b,Balasis2011chap}. We also investigated the existence of a dynamical analogy between a single epileptic seizure / earthquake generation by means of scale-free statistics, namely, the traditional Gutenberg-Richter distribution of event sizes and the distribution of the waiting time until the next event. Our study offers and justifies a positive answer. One universal footprint seen in complex systems is self-affinity. The results reveal that the populations of (a) earthquakes occurred in different faults, (b) epileptic seizures occurred in different patients, (c) fracto-EM-pulses rooted in the fracture of the backbone of strong entities distributed along a single fault sustaining the system, and (d) electric pulses included in a single epileptic seizure follow the same scale-free statistics. Importantly, the analogy is extended up to the laboratory scale of fracture. 

It has reported that electromagnetic flashes of low-energy $\gamma-rays$ emitted during multi-fracturing on a neutron star, electromagnetic pulses emitted in the laboratory by a disordered material subjected to an increasing external load, and electromagnetic pulses emitted in the field during the activation of a fault share distinctive statistical properties with earthquakes, such as power-law energy distributions \cite{Cheng1996,Kossobokov2000,Rabinovitch2001,Sornette2002,Kapiris2004,Telesca_7}. The neutron starquakes may release strain energies up to 1046 erg, while, the fractures in laboratory samples release strain energies approximately a fraction of an erg. An earthquake fault region can build up strain energy up to approximately 1026 erg for the strongest earthquakes. Consequently, the limits of similarity in the dynamics of different multi-fracturing extreme events are dramatically expanded.

In summary, our experimental results support previously reported theoretical and experimental arguments \cite{Hopfield1994,Herz1995,Rundle2002} that dynamics of earthquakes and neurodynamics can be analyzed within similar mathematical frameworks, while the dynamical analogy by means of non-extensive and traditional free scale laws between earthquakes and epileptic seizures is extended up to solar flares, magnetic storms and neutron starquakes dynamics. 

\section*{Acknowledgments}
The second author (G.M) would like to acknowledge research funding received by the Greek State Scholarships Foundation (IKY).

\bibliography{quakesofbrain}

\end{document}